\documentclass[twocolumn]{autart}

\usepackage{amsmath,amssymb,pstricks,color,dsfont}
\usepackage{natbib}
\usepackage{graphicx}
\usepackage{psfrag,graphicx,epsfig}
\usepackage{algorithm,algorithmic,xspace}
\usepackage{subfigure}
\usepackage{float}
\usepackage{graphicx}
\usepackage{epstopdf}
\usepackage{savesym}
\usepackage{multirow}
\savesymbol{AND}
\allowdisplaybreaks

\newcommand{\bremark}{\begin{remark} \begin{rm} }
\newcommand{\eremark}{ \end{rm} \rule{1mm}{2mm}
\end{remark} }
\newcommand{\btheorem}{\begin{theorem} \begin{rm} }
\newcommand{\etheorem}{ \end{rm} \rule{1mm}{2mm}
\end{theorem} }
\newcommand{\blemma}{\begin{lemma} \begin{rm} }
\newcommand{\elemma}{ \end{rm} \rule{1mm}{2mm}
\end{lemma} }
\newcommand{\bcorollary}{\begin{corollary} \begin{rm} }
\newcommand{\ecorollary}{ \end{rm} \rule{1mm}{2mm}
\end{corollary} }
\newcommand{\bdefinition}{\begin{definition}\begin{rm} }
\newcommand{\edefinition}{ \end{rm} \rule{1mm}{2mm}
\end{definition} }
\newcommand{\bproposition}{\begin{proposition} \begin{rm} }
\newcommand{\eproposition}{ \end{rm} \rule{1mm}{2mm}
\end{proposition} }
\newcommand{\bexample}{\begin{example} \begin{rm} }
\newcommand{\eexample}{ \end{rm} \rule{1mm}{2mm}
\end{example} }
\newcommand{\basm}{\begin{assumption} \begin{rm}}
\newcommand{\easm}{\end{rm} \rule{1mm}{2mm}
\end{assumption}}

\newtheorem{theorem}{\bf Theorem}[section]
\newtheorem{lemma}{\bf Lemma}[section]
\newtheorem{definition}{\bf Definition}[section]
\newtheorem{remark}{\bf Remark}[section]
\newtheorem{corollary}{\bf Corollary}[section]
\newtheorem{proposition}{\bf Proposition}[section]
\newtheorem{example}{\bf Example}[section]
\newtheorem{assumption}{\bf Assumption}[section]
\newcommand\oprocendsymbol{\hbox{$\bullet$}}
\newcommand\oprocend{\relax\ifmmode\else\unskip\hfill\fi\oprocendsymbol}

\date{}

\begin{document}

\begin{frontmatter}

\title{Privacy preserving distributed optimization using homomorphic encryption\thanksref{footnoteinfo}}

\thanks[footnoteinfo]{This work was partially supported by NSA grant H98230-15-1-0289 and NSF grant CNS-1505664.}

\author[PSU]{Yang Lu}\ead{yml5046@psu.edu},
\author[PSU]{Minghui Zhu}\ead{muz16@psu.edu}

\address[PSU]{School of Electrical Engineering and Computer Science, Pennsylvania State University, 201 Old Main, University Park, PA, 16802 USA}


\begin{abstract}
This paper studies how a system operator and a set of agents securely execute a distributed projected gradient-based algorithm. In particular, each participant holds a set of problem coefficients and/or states whose values are private to the data owner. The concerned problem raises two questions: how to securely compute given functions; and which functions should be computed in the first place. For the first question, by using the techniques of homomorphic encryption, we propose novel algorithms which can achieve secure multiparty computation with perfect correctness. For the second question, we identify a class of functions which can be securely computed. The correctness and computational efficiency of the proposed algorithms are verified by two case studies of power systems, one on a demand response problem and the other on an optimal power flow problem.
\end{abstract}

\end{frontmatter}

\section{Introduction\label{section one}}

In the last decades, distributed optimization has been extensively studied and broadly applied to coordinate large-scale networked systems \cite{Bertsekas.Tsitsiklis:97}, \cite{MZ-SM-distributed-book:2015}, \cite{fb-jc-sm:09}. In distributed optimization, the participants collectively achieve network-wide goals via certain mechanisms driven by data sharing between the participants. Such data sharing, however, causes the concern that private information of legitimate entities could be leaked to unauthorized ones. Hence, there is a demand to develop new distributed optimization algorithms that can achieve network-wide goals and simultaneously protect the privacy of legitimate entities.

The problem of interest is closely relevant to the one where a group of participants aim to compute certain functions over their distributed private inputs such that each participant's inputs remain private after the computation. To protect privacy, two questions need to be answered \cite{YL-BP:09}: (i) How to securely compute given functions with distributed private inputs so that the computation process does not reveal anything beyond the function outputs? This question is referred to as secure multiparty computation (SMC) \cite{Hazay}, \cite{RC-ID-JBN:2015}. (ii) Which functions should be computed in the first place so that the adversary cannot infer private inputs of benign participants from function outputs? In this paper, we refer to this question as input-output inference (IOI).

\emph{Literature review on homomorphic encryption}. For the question of SMC, homomorphic encryption is a powerful tool, and has been applied to various problems including, e.g., statistical analysis \cite{ES-THHC-EGR-RC-DS:2011} and data classification \cite{ZY-SZ-RNW:2005}. It is because homomorphic encryption allows certain algebraic operations to be carried out on ciphertexts, thus generating an encrypted result which, when decrypted, matches the result of operations performed on plaintexts. A detailed literature review on homomorphic encryption is provided in Section \ref{literature review on HE}. It is worthy to mention that private key fully (and surely partially) homomorphic encryption schemes and public key partially homomorphic encryption schemes can be efficiently implemented (see Section \ref{literature review on HE}). Recently, homomorphic encryption has been applied to control and optimization problems, e.g., potential games \cite{YL-MZ:2015}, quadratic programs \cite{YS-KG-AA-GJP-SAS-MS-PT:2016}, and linear control systems \cite{KK-TF:2015}, \cite{FF-IS-NB:2017}. Most existing homomorphic encryption schemes only work for binary or non-negative integers. In contrast, states and parameters in most distributed optimization problems are signed real numbers.

\emph{Literature review on IOI}. SMC alone is not enough for data privacy. Even if a computing scheme does not reveal anything beyond the outputs of the functions, the function outputs themselves could tell much information about the private inputs. This raises the second question of IOI. A representative set of papers, e.g., \cite{MRC-ACM-FBS:2009}, \cite{PM-SM-MH-MS:2011}, \cite{PM-MH-JK-MS:2012}, adopted the method of belief set. In particular, each participant maintains a belief which is an estimate of the other participant's knowledge. When the belief is below certain threshold, the participant continues computation. Otherwise, the participant rejects the computation. This approach is only applicable to discrete-valued problems with constant private inputs. To the best of our knowledge, the issue of IOI in real-valued time series, e.g., generated by dynamic systems, has been rarely studied.

\emph{Contributions}. This paper investigates how a system operator and a set of agents execute a distributed gradient-based algorithm in a privacy preserving manner. In particular, on the one hand, each agent's state and feasible set are private to itself and should not be disclosed to any other agent or the system operator; on the other hand, each of the agents and the system operator holds a set of private coefficients of the component functions and each coefficient should not be disclosed to any participant who does not initially hold it. The privacy issue of the gradient-based algorithm is decomposed into SMC and IOI.

For the question of SMC, we first propose a private key fully homomorphic encryption scheme to address the case where jointly computed functions are arbitrary polynomials and the system operator could only launch temporarily independent attacks, i.e., at each step of the gradient-based algorithm, the system operator only uses the data received at the current step to infer the agents' current states, but does not use previous data to collectively infer the agents' states in the past. We notice that the assumption of temporarily independent attacks is crucial for the private key homomorphic encryption setting and the security might be compromised if this assumption does not hold. Similar assumptions have been widely used in database privacy \cite{RB-AB-VG-SL-AT:2011}, \cite{CD-MN-TP-GNR:2010}, \cite{TZ-GL-WZ-PSY:2017}. Please refer to Remark \ref{remark one shot attack} for the detailed discussion on this assumption. To deal with real numbers, we propose mechanisms for transformations between real numbers and integers as pre and post steps of plain homomorphic encryption schemes for integers, and provide the condition that the key should satisfy to guarantee the correctness of decryption and post-transformation. The proposed technique to handle real numbers can be applied to arbitrary homomorphic encryption schemes. All the agents encrypt their states and coefficients by a key which is unknown to the system operator. The system operator computes the polynomial functions with the encrypted data and sends the results to the target agent that initiates the computation. The target agent can then perform decryption by the same key. We prove that the algorithm correctly computes the functions and meanwhile prove by the simulation paradigm that each agent does not know anything beyond what it must know (i.e., its inputs and the outputs of its desirable functions) and it is computationally hard for the system operator to infer the private data of the agents.

We then propose a public key partially homomorphic encryption computing algorithm to address the case where jointly computed functions are affine and the system operator could launch causal attacks. The secure computation of affine functions is carried out by the Paillier encryption scheme. Similar techniques are applied to handle real numbers. We prove that the proposed algorithm computes the correct values of the functions and after the computation, each participant does not learn anything beyond what it must know and it is computationally hard for the system operator to distinguish the private data of the agents.

For the question of IOI, we provide a control-aware definition which, informally, states that a function is secure to compute if, given the output of the function, the adversary cannot uniquely determine the private inputs of any participant and meanwhile the uncertainty about the inputs is infinite. The definition is inspired by the notion of observability in control theory and is consistent with the uncertainty based privacy notions in database analysis \cite{DGM:1996}, \cite{AB-CF-SJ:2000}. For a class of quadratic functions, we derive sufficient conditions of IOI by exploiting the null vectors of the coefficient weight matrices and the constant terms.

The correctness and computational efficiency are verified by two case studies of power systems, one on a demand response problem and the other on an optimal power flow problem.

In our earlier paper \cite{YL-MZ:2015}, homomorphic encryption was applied to discrete potential games. In \cite{YL-MZ:2015}, the issue of IOI was not discussed and the proofs were omitted. The current paper studies a class of distributed algorithms which have been widely used to solve distributed optimization, convex games and stochastic approximation. In the current paper, both the issues of SMC and IOI are studied with formal proofs. For the issue of SMC, the current paper adopts two different homomorphic encryption schemes which are based on some computationally hard problems and achieve higher privacy level than that of \cite{YL-MZ:2015}.

\emph{Literature review on other secure computation techniques}. Besides homomorphic encryption, several other techniques have been adopted to address privacy issues in control and optimization.

The first branch of works uses differential privacy \cite{CD:06}, \cite{Dwork} as the security notion, e.g., \cite{MTH-ME:2015}, \cite{ZH-SM:2012}, \cite{JLN-GJP:2014}, \cite{SH-UT-GJP:2017} developed differentially private algorithms for distributed optimization, consensus and filtering problems, respectively. Differentially private schemes add persistent randomized perturbations into data to protect privacy \cite{QG-PV:2014}. For control systems, such persistent perturbations could potentially deteriorate system performance. In contrast, homomorphic encryption schemes do not introduce any perturbation and are able to find exact solutions.

The second branch of works uses obfuscation techniques \cite{JD-FK:2011} to protect coefficient confidentiality for optimization problems in cloud computing. Related works include, e.g., \cite{ARB-DKM-BCL-PR:2013}, \cite{ARB-DKM-PR-BCL:2012}, \cite{CW-KR-JW:2016}, which studied optimal power flow and linear programming problems, respectively. Existing obfuscation techniques are only applicable to centralized optimization problems with linear or quadratic cost functions in order to have the property that inverting the linear obfuscation transformation returns the optimal solution of the original problem.

The third branch of works adopts the techniques of secret sharing, e.g., in our earlier work \cite{YL-MZ:NECSYS2015Privacy}, the Shamir's secret sharing scheme \cite{Shamir} was used to achieve state privacy for distributed optimization problems on tree topologies. In \cite{YL-MZ:NECSYS2015Privacy}, the functions are restricted to be linear and the privacy of coefficients is not taken into account.

\emph{Notations}. The vector $\textbf{1}_n$ denotes the column vector with $n$ ones. Given a finite index set $\Omega$, let $[A_i]_{i\in \Omega}$ denote the column-wise stack of $A_i$ for all $i\in \Omega$, where $A_i$'s are matrices with the same number of columns. When there is no confusion in the context, we drop the subscript $i\in\Omega$ and use $[A_i]$. Given matrices $A_1,\cdots,A_N$ with the same column number, let $A_{-i}$ denote $[A_j]_{j\neq i}$. Given matrices $A_1,\cdots,A_N$, denote by ${\rm diag}\{A_1,\cdots,A_N\}$ the block diagonal matrix for which the sub-matrix on the $i$-th diagonal block is $A_i$ and all the off-diagonal blocks are zero matrices. For a matrix or vector $A$, $\|A\|$ denotes the 2-norm of $A$, and $\|A\|_{\rm max}$ denotes the max norm of $A$, i.e., the absolute value of the element of $A$ with the largest absolute value. For any positive integer $n$, $\mathbb{R}^n$, $\mathbb{R}_{\geq0}^n$, $\mathbb{R}_{>0}^n$, $\mathbb{Z}^n$, $\mathbb{Z}_{\geq0}^n$ and $\mathbb{Z}_{>0}^n$ denote the sets of real, non-negative real, positive real, integer, non-negative integer, and positive integer column vectors of size $n$, respectively. Let $\mathbb{N}$ denote the set of natural numbers. Given $w\in\mathbb{Z}_{>0}$, $\mathbb{Z}_w \buildrel \Delta \over = \{0,1,\cdots,w-1\}$ and $\mathbb{Z}_w^*$ denotes the set of positive integers which are smaller than $w$ and do not have common factors other than $1$ with $w$. Given a finite set $C$, denote by $|C|$ its cardinality. Given a non-empty, closed and convex set $Z\subseteq \mathbb{R}^m$, $\mathbb{P}_Z$ denotes the projection operation onto $Z$, i.e., for any $z\in\mathbb{R}^m$, $\mathbb{P}_Z[z]={\rm argmin}_{y\in Z}\|y-z\|$. The operator $mod$ denotes the modulo operation such that, given any $a,b\in\mathbb{Z}$ and $b$ non-zero, $a\mod b$ returns the remainder of $a$ modulo $b$. Given any $p,q\in\mathbb{Z}_{>0}$, ${\rm gcd}(p,q)$ and ${\rm lcm}(p,q)$ denote the greatest common divisor and the least common multiple of $p$ and $q$, respectively. Given a polynomial function $f$, ${\rm deg}(f)$ denotes the degree of $f$.

\section{Problem formulation}

In this section, we first formulate the problem of gradient-based distributed optimization and its privacy issues. After that, we discuss the attacker model and privacy notions. Finally, we propose a mechanism for transformation between integers and real numbers.

\subsection{Gradient-based distributed optimization\label{section SCP formulation}}

Consider a network comprising a set of agents $\mathcal{V} = \{1,\cdots,N\}$ and a system operator ($\rm SO$). Each agent $i$ has a state $x_i=[x_{i\ell}]\in X_i\subseteq\mathbb{R}^{n_i}$, where $X_i$ is the feasible set of $x_i$. Let $x =[x_i]_{i\in\mathcal{V}}\in\mathbb{R}^n$ with $n =\sum_{\ell=1}^Nn_i$. Throughout the paper, we have the following assumption on the communication topology.

\begin{assumption}
For each $i\in\mathcal{V}$, there is an undirected communication link between agent $i$ and the system operator.\hfill\rule{1mm}{2mm}
\label{asm: star topology}
\end{assumption}

Many networked systems naturally admit a system operator, e.g., Internet, power grid and transportation systems \cite{Wu.Moslehi.Bose:89}, \cite{MC-SHL-ARC-JCD:07a}. Assumption \ref{asm: star topology} is widely used in, e.g., network flow control \cite{LL-99} and demand response \cite{NL-LC-SHL:2011}.

The agents aim to address a distributed optimization problem by the projected gradient method, i.e., each agent $i$ updates its state as follows:
\begin{align}
\label{SCP of DC}
x_i(k+1)=\mathbb{P}_{X_i}[x_i(k)-\gamma(k)\Phi_i(x(k))].
\end{align}
In \eqref{SCP of DC}, $k$ denotes the discrete step index; $\gamma(k)>0$ is the step size at step $k$; $\Phi_i:\mathbb{R}^{n}\to\mathbb{R}^{n_i}$ is the first-order gradient of certain functions with respect to $x_i$. Denote the $\ell$-th element of $\Phi_i$ by $\Phi_{i\ell}$. Notice that $\Phi_i$ in general depends on the whole $x$. The agents update their states by \eqref{SCP of DC} at each step $k$ and aim to achieve convergence of a solution of the distributed optimization problem.

\begin{remark}
Update rule \eqref{SCP of DC} has been widely used to solve distributed optimization, convex games, variational inequalities and stochastic approximation in, e.g., \cite{FF-JSP:03}, \cite{HJK-GGY:1997}, \cite{MZ-EF:Auto14}, \cite{Bertsekas.Tsitsiklis:97} and the references therein. The convergence of update rule \eqref{SCP of DC} is well studied in these references. To guarantee convergence, convexity is usually assumed \cite{Bertsekas.Tsitsiklis:97}, \cite{Bertsekas:09}. However, these references do not consider the privacy issues (identified next). In the current paper, we assume that the underlying distributed optimization problem satisfies existing conditions for convergence and do not provide any analysis on convergence issues. Instead, the current paper solely focuses on the problem of how to carry out \eqref{SCP of DC} in a privacy preserving manner.\hfill\rule{1mm}{2mm}
\label{remark: PGM}
\end{remark}

Throughout the paper, we have the following polynomial function assumption.

\begin{assumption}
For each $i\in\mathcal{V}$ and each $\ell\in\{1,\cdots,n_i\}$, function $\Phi_{i\ell}$ is a polynomial of $x$.\hfill\rule{1mm}{2mm}
\label{asm: polynomial}
\end{assumption}

\begin{remark}
By \cite{RAD-GGL:1993}, the algebra of polynomials can approximate any continuous function over a compact domain arbitrarily well. We refer to \cite{RAD-GGL:1993} for the details of polynomial approximation. In this paper, we do not consider division operations as divisions largely complicate the task of secure computing algorithm design because existing homomorphic encryption schemes do not directly support division operations \cite{NF-PP:2016}.\hfill\rule{1mm}{2mm}
\label{remark: polynomial}
\end{remark}

\subsection{Privacy issues\label{section privacy issues}}

In this subsection, we identify the privacy issues in the execution of \eqref{SCP of DC}. Denote by $C_\Phi$ the set of coefficients of the functions $(\Phi_1,\cdots,\Phi_N)$. For each $i\in\mathcal{V}\cup\{{\rm SO}\}$, participant $i$ holds a subset of coefficients of $C_\Phi$, denoted by $C_\Phi^i$. It holds that $\bigcup_{i\in\mathcal{V}\cup\{{\rm SO}\}}C_\Phi^i=C_\Phi$. Notice that a coefficient could be shared by multiple participants.

For each $i\in\mathcal{V}\cup\{{\rm SO}\}$, the private data of participant $i$ includes the following three parts:
\begin{itemize}
\item The sequence of its states $\{x_i(k)\}$ (only for $i\in\mathcal{V}$).
\item Its feasible set $X_i$ (only for $i\in\mathcal{V}$).
\item The set $C_\Phi^i$ of coefficients it holds (for $i\in\mathcal{V}\cup\{{\rm SO}\}$).
\end{itemize}

For each $i\in\mathcal{V}$, its state sequence $\{x_i(k)\}$ and feasible set $X_i$ should not be disclosed to any other participant $j\in(\mathcal{V}\cup\{{\rm SO}\})\backslash\{i\}$. In distributed optimization, the state and feasible set of an agent could expose much information about the agent's behaviors and thus should not be leaked to other entities. For example, a demand response problem involves a set of power customers (agents) and a utility company (system operator). Each customer aims to achieve the optimal power load such that the total cost induced by disutility and load charge is minimized and the load benefit is maximized. This problem can be formulated as a distributed optimization problem in which each customer's state is its power load \cite{NL-LC-SHL:2011}. It has been shown that power load profiles at a granularity of 15 minutes may reveal whether a child is left alone at home and at a finer granularity may reveal the daily routines of customers \cite{YG-YC-YG-YF:2016}. Moreover, each coefficient $r\in C_\Phi$ should not be disclosed to any participant $i\in\mathcal{V}\cup\{\rm SO\}$ such that $r\notin C_\Phi^i$. The coefficients of a distributed optimization problem are usually related to system parameters and, in some scenarios, it is crucial to keep the system parameters private to unauthorized entities. For example, an optimal power flow problem involves a set of power generators (agents) and an energy manager (system operator). Each generator aims to find its optimal mechanical power and phase angle such that the operating cost is minimized. This problem can be formulated as a distributed optimization problem in which the constraints are affine and the coefficients of the constraints are the line-dependent parameters of the power system \cite{Wood:1996}. It has been pointed out that leakage of line-dependent parameters could be financially damaging and even cause potential threat to national security \cite{ARB-DKM-BCL-PR:2013}, \cite{ARB-DKM-PR-BCL:2012}.

For each $i\in\mathcal{V}$, we assume that the system operator knows the structure of $\Phi_i$ but is unaware of the values of $x$ and the coefficients other than $C_\Phi^{\rm SO}$. For each step $k$, we assume that the step size $\gamma(k)$ is identical for all the agents and known to all the agents.

\begin{figure}
\begin{center}
\includegraphics[width=0.6\linewidth]{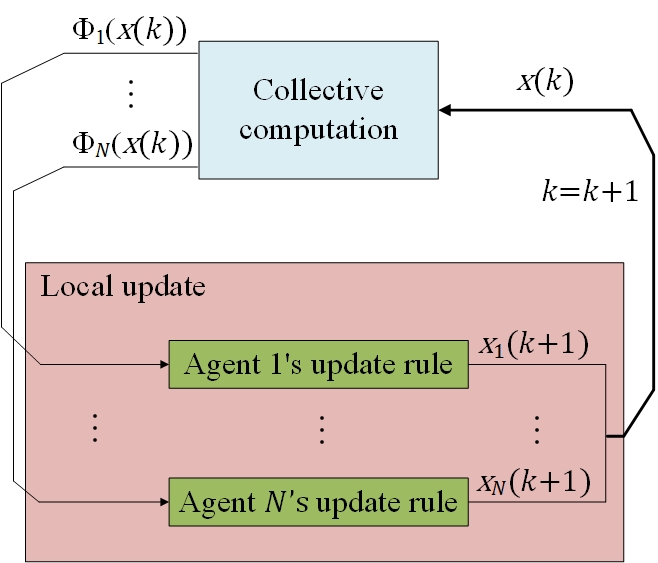}
\caption{Block diagram of the computation problem \eqref{SCP of DC}}
\label{algorithm decomposition}
\end{center}
\end{figure}

With the privacy issue demonstrated above, it would be helpful to highlight our secure computation problem by decomposing the execution of \eqref{SCP of DC} into two parts: the collective computation part where all participants collectively compute $\Phi_i(x(k))$'s and the local update part where each agent $i$ updates its state $x_i(k)$; see Fig. \ref{algorithm decomposition}. In \eqref{SCP of DC}, for each agent $i$ to update $x_i(k)$, it needs the value of $\Phi_i(x(k))$, whose computation depends on private data of other participants and thus requires data exchange between the participants. Once agent $i$ obtains the value of $\Phi_i(x(k))$, it can locally update $x_i(k)$ by \eqref{SCP of DC}. In this two-part process, the privacy issue raises two questions:
\begin{itemize}
\item Secure multiparty computation (SMC): How to securely compute $\{\Phi_i(x(k))\}$ for each agent $i\in\mathcal{V}$ such that agent $i$ knows the correct values of these functions and meanwhile no agent can learn anything beyond its function outputs?
\item Input-output inference (IOI): Which functions $(\Phi_1,\cdots,\Phi_N)$ should be computed such that, for each $i\in\mathcal{V}$, for any $K\in\mathbb{N}$, by receiving the sequence $\{\Phi_i(0),\cdots,\Phi_i(K)\}$, agent $i$ cannot infer the private data of other participants?
\end{itemize}

\subsection{Attacker model and privacy notions\label{security issues}}

In this subsection, we identify the attacker model and privacy notions adopted in this paper.

In this paper, we are concerned with semi-honest adversaries, i.e., any adversary $i\in\mathcal{V}\cup\{\rm SO\}$ correctly follows the algorithm but attempts to use its received messages throughout the execution of the algorithm to infer other participants' private data (\cite{Hazay}, pp-20). Semi-honest adversary model is broadly used in SMC \cite{RC-ID-JBN:2015}, \cite{Hazay} and has been adopted in various applications, e.g., linear programming, dataset process and consensus \cite{JD-FK:2011}, \cite{MJF-KN-BP}, \cite{ZH-SM:2012}. Moreover, the system operator is able to eavesdrop all the communication links while we assume that any agent cannot eavesdrop the communication links between other participants. We also assume that the adversaries do not collaborate to infer the information of benign participants. Instead, if multiple adversaries collaborate, they are viewed as a single adversary.

We next introduce the privacy notions adopted in this paper.

\emph{$\bullet$ Privacy notion for SMC.} We adopt the standard language of \emph{view} in cryptography to develop the notion.

Consider a set of $M$ parties where each party $i$ holds a set of private data, denoted by $y_i\in\mathbb{R}^{d_i}$. Let $y =[y_i]\in\mathbb{R}^d$ with $d =\sum_{i=1}^Md_i$. Each party $i$ aims to compute a set of functions $h_i:\mathbb{R}^d\to\mathbb{R}^{r_i}$ which depends on the overall $y$. Let $h =[h_i]:\mathbb{R}^d\to\mathbb{R}^r$ with $r =\sum_{i=1}^Mr_i$. Let $\Pi$ be an algorithm that enables the $M$ parties to collectively compute $h$. For each party $i$, its input for $\Pi$ is $y_i$. Let $t_i$ be the total number of messages received by party $i$ during the execution of $\Pi$ and denote these messages by $v_1^i,\cdots,v_{t_i}^i$. Given $y$, the view of party $i$ during the execution of $\Pi$, denoted by ${\rm VIEW}_i^\Pi(y)$, is defined as ${\rm VIEW}_i^\Pi(y)\buildrel \Delta \over =\{y_i,v_1^i,\cdots,v_{t_i}^i\}$. To introduce the definition of privacy for SMC, we first need to introduce the definition of computational indistinguishability.

\begin{definition}[\cite{RC-ID-JBN:2015}]%, \cite{YL-BP:09}]
\label{def indistinguishable}
Let $\mathcal{X}_{[\kappa]}=\{\mathcal{X}_1,\cdots,\mathcal{X}_\kappa\}$ and $\mathcal{Y}_{[\kappa]}=\{\mathcal{Y}_1,\cdots,\mathcal{Y}_\kappa\}$ be two distribution ensembles, where for each $\ell\in\{1,\cdots,\kappa\}$, $\mathcal{X}_\ell$ and $\mathcal{Y}_\ell$ are random variables with range $R_\ell$. We say that $\mathcal{X}_{[\kappa]}$ and $\mathcal{Y}_{[\kappa]}$ are computationally indistinguishable, denoted $X\mathop  \equiv \limits^c Y$, if for every non-uniform probabilistic polynomial-time distinguisher $D$, every positive polynomial $p:\mathbb{N}\to\mathbb{R}_{>0}$, and every sufficiently large $\kappa\in\mathbb{N}$, the following holds:
\begin{align*}
|\Pr[D(\mu)=1]-\Pr[D(\nu)=1]|<{1}/{p(\kappa)}
\end{align*}
where $\mu$ and $\nu$ are random samples drawn from $\mathcal{X}_{[\kappa]}$ and $\mathcal{Y}_{[\kappa]}$, respectively.
\hfill\rule{1mm}{2mm}
\end{definition}

The following privacy notion for SMC is standard in the literature.

\begin{definition}[\cite{RC-ID-JBN:2015}]%, \cite{YL-BP:09}]
\label{secure compute def}
Let $\Pi$ be an algorithm for computing $h$. We say that $\Pi$ securely computes $h$ if there exists a probabilistic polynomial-time algorithm $S$ such that for each party $i\in\{1,\cdots,M\}$ and every possible $y$, it holds that $S(i,y_i,h_i(y))\mathop  \equiv \limits^c{\rm VIEW}_i^{\Pi}(y)$, where $S(\cdot)$ denotes the overall messages that can be seen after the execution of $S$.
\hfill\rule{1mm}{2mm}
\end{definition}

Definition \ref{secure compute def} states that whatever a party receives during the computing process can be constructed only via its own inputs and the outputs of the functions to compute. In other words, the computing process reveals nothing more than what a party has to learn.

\emph{Encryption privacy.} In this paper, we use homomorphic encryption to tackle the issue of SMC for the agents. This technique requires some computing entity, which could be semi-honest, to perform function evaluations over encrypted data. In this paper, the system operator is used as the computing entity. Roughly speaking, the agents encrypt their private data and send the encrypted data to the system operator. The system operator evaluates the agents' functions $\Phi_i$'s over the received encrypted data and sends the results to the agents. The agents then decrypt the results to obtain the true values of $\Phi_i$'s.

On the one hand, the above homomorphic encryption setting could tackle the SMC issue as the agents only receive aggregate function outputs, but do not receive any form of individual private data. On the other hand, since the system operator could be semi-honest, we need to ensure encryption privacy against the system operator, that is, the system operator cannot infer the private data of the agents from the encrypted data it receives. In the homomorphic encryption literature \cite{XY-RP-EB:2014}, \cite{CP-JP:2010}, \cite{WS:2011}, two different notions have been widely used to claim or define encryption privacy against the computing entity: (i) plaintexts not efficiently solvable \cite{RLR-AS-LA:1978} and (ii) plaintexts not efficiently distinguishable \cite{SG-SM:1982}. This paper adopts both the two notions, illustrated next.

(i) Plaintexts not efficiently solvable. By this notion, an encryption scheme is secure if the adversaries cannot solve plaintexts via observing ciphertexts by any polynomial-time algorithms. For an encryption scheme that adopts this privacy notion, breaking the scheme (i.e., solving plaintexts) is or is believed to be computationally hard. However, in many cases, it is not proved that the leveraged hard problem itself indeed does not admit a polynomial-time solving algorithm, but rather people have not found one yet. Hence, for this privacy notion, the security level of an encryption scheme is usually not established by a proof, but claimed in a heuristic manner by checking all existing solving algorithms and claiming that none of them is efficient (usually given that the key length is large enough). Well-known encryption schemes that adopt this notion of security and claim their security in this heuristic manner include public key encryption scheme RSA and private key encryption schemes DES (Data Encryption Standard) and AES (Advanced Encryption Standard).

(ii) Plaintexts not efficiently distinguishable. Roughly speaking, an encryption scheme is secure in this sense if the adversaries cannot distinguish between any plaintexts via observing ciphertexts by any polynomial-time algorithms. For the term indistinguishability, informally, it means that given an encryption of a message randomly chosen from any two-element message space determined by the adversary, the adversary cannot identify the message choice with probability significantly better than $0.5$. It is clear that this privacy notion is stronger than the above one, since indistinguishability surely implies unsolvability, but the reverse direction may not be true (it could be hard to find which is the plaintext but easy to tell which is not). A standard definition for this privacy notion is semantic security, defined next.

\begin{definition}[\cite{RC-ID-JBN:2015}]%, \cite{YL-BP:09}]
\label{def encryption semantic security}
Let $E$ be an encryption scheme which outputs ciphertext $E(x)$ on plaintext $x$. An adversary chooses any plaintexts $x,x'$ and asks for their ciphertexts from the scheme holder. The scheme holder outputs ciphertexts $E(x)$ and $E(x')$ and sends them to the adversary without telling the adversary which ciphertext corresponds to which plaintext. We say that $E$ is semantically secure if for any plaintexts $x,x'$ chosen by the adversary in the above setting, it holds that $E(x)\mathop  \equiv \limits^cE(x')$ to the adversary.
\hfill\rule{1mm}{2mm}
\end{definition}

Definition \ref{def encryption semantic security} renders semantically secure encryption schemes provable. Usually, the semantic security of an encryption scheme is established upon an assumption that certain problem is hard to solve, and then the proof of semantic security is performed by reducing the distinguish task to solving the concerned hard problem. Well-known semantically secure encryption schemes include public key encryption schemes Goldwasser-Micali, ElGamal, Paillier and Boneh-Goh-Nissim.

\emph{$\bullet$ Privacy notions for IOI.} To the best of our knowledge, there is no dominating definition for the question of IOI in cryptography community. In this paper, we adopt the following definition to quantify the uncertainty an adversary has about the private inputs of other parties given the outputs of the functions. The definition is inspired by the notion of observability in control theory and is consistent with the uncertainty based privacy notions in the field of database analysis \cite{DGM:1996}, \cite{AB-CF-SJ:2000}.

\begin{definition}
\label{secure from output}
Denote the true valuation of $y_i$ by $\bar y_i$ for each party $i$ and denote $\bar y=[\bar y_i]$. For each party $i$, for each $j\neq i$, denote $\Delta_j^i(\bar y)=\{y_j\in\mathbb{R}^{d_j}:\exists y_\ell\in\mathbb{R}^{d_\ell}\;{\rm for\;all\;}\ell\neq i,j\;{\rm s.t.}\;h_i(\bar y_i,y_j,[y_\ell]_{\ell\neq i,j})=h_i(\bar y)\}$, and denote ${\rm Dist}_j^i(\bar y)=\sup_{y_j,y_j'\in\Delta_j^i(\bar y)}\|y_j-y_j'\|_2$. Given $\bar y$, for each $j\in\{1,\cdots,M\}$, the amount of uncertainty in $y_j$ in the input-output inference of $h$ is defined as $\min_{i\neq j}{\rm Dist}_j^i(\bar y)$. We say that the function $h$ resists input-output inference with unbounded uncertainty if the amount of uncertainty in $y_j$ is infinite for all $j\in\{1,\cdots,M\}$ and for any $\bar y\in\mathbb{R}^d$.\hfill\rule{1mm}{2mm}
\end{definition}

In the subsequent sections, we first study the question of SMC in Sections \ref{private key section} and \ref{public key section}. After that, we analyze the question of IOI in Section \ref{observability section}.

\subsection{Transformation between integers and real numbers\label{section integer real transformation}}

In distributed optimization problems, the variables and coefficients are usually real numbers. However, most existing homomorphic encryption schemes rely on certain modular operations and only work for integers. Therefore, we need a mechanism for transformation between real numbers and integers. Throughout the paper, the accuracy level is set by a parameter $\sigma\in\mathbb{N}$, which means that for any real number, $\sigma$ decimal fraction digits are kept while the remaining decimal fraction digits are dropped. We assume that the value of $\sigma$ is known to the system operator and all the agents. For a real number $r$ with $\sigma$ decimal fraction digits, it is transformed into an integer $z_r$ simply by $z_r=10^\sigma r$. The following function is needed to transform a non-negative integer into a signed real number. Given $\sigma\in\mathbb{N}$ and an odd positive integer $w$, an integer $z\in \mathbb{Z}_w$ is transformed into a signed real number with $\sigma$ decimal fraction digits by the following function parameterized by $\sigma$ and $w$:
\begin{align}
\label{integer to real}
T_{\sigma,w}(z) = \left\{ {\begin{array}{*{20}{l}}
{{z}/{10^\sigma},\,{\rm if}\,0 \le z \le {{(w - 1)}}/{2}},\\
{{(z-w)}/{10^\sigma},\,{\rm if}\, {{(w + 1)}}/{2} \le z < w}.
\end{array}} \right.
\end{align}

The following property holds for the $T$ function \eqref{integer to real} and is crucial to guarantee correctness of computations over signed real numbers.

\begin{lemma}
\label{T function property}
Given an odd positive integer $w$, for any $r\in\mathbb{R}$ with $\sigma$ decimal fraction digits such that $|10^\sigma r|\leq(w-1)/2$, it holds that $T_{\sigma,w}(10^\sigma r\mod w)=r$.\hfill\rule{1mm}{2mm}
\end{lemma}

\begin{pf} Since $|10^\sigma r|\leq (w-1)/2$, it is either $0\leq10^\sigma r\leq(w-1)/2$ or $-(w-1)/2\leq10^\sigma r<0$.

Case I: $0\leq10^\sigma r\leq(w-1)/2$. For this case, we have $10^\sigma r\mod w=10^\sigma r$. Hence, by \eqref{integer to real}, we have $T_{\sigma,w}(10^\sigma r\mod w)=T_{\sigma,w}(10^\sigma r)=(10^\sigma r)/10^\sigma=r$.

Case II: $-{(w-1)}/{2}\leq10^\sigma r<0$. For this case, we have $10^\sigma r\mod w=w+10^\sigma r$. It is clear that $(w+1)/2\leq w+10^\sigma r<w$. Hence, by \eqref{integer to real}, we have $T_{\sigma,w}(10^\sigma r\mod w)=T_{\sigma,w}(w+10^\sigma r)=(w+10^\sigma r-w)/10^\sigma=r$.
\end{pf}

\section{Private key secure computation algorithm\label{private key section}}

In this section, we propose a secure computation algorithm for \eqref{SCP of DC} which is based on a private key fully homomorphic encryption scheme.

\subsection{Preliminaries\label{section polynomial preliminaries}}

First, each participant $i\in\mathcal{V}\cup\{\rm SO\}$ constructs a set $\tilde C_\Phi^i$ of coefficients such that $(\tilde C_\Phi^1,\cdots,\tilde C_\Phi^N,\tilde C_\Phi^{\rm SO})$ have the following properties: (i) $\tilde C_\Phi^{\rm SO}=C_\Phi^{\rm SO}$; (ii) $\tilde C_\Phi^i\subseteq C_\Phi^i$ for all $i\in\mathcal{V}$; (iii) the sets $\tilde C_\Phi^i$'s form a partition of $C_\Phi$, i.e., $\bigcup_{i\in\mathcal{V}\cup\{\rm SO\}}\tilde C_\Phi^i=C_\Phi$ and $\tilde C_\Phi^i\cap\tilde C_\Phi^j=\emptyset$ for any $i,j\in\mathcal{V}\cup\{\rm SO\}$ and $i\neq j$. Such a construction can be realized by a number of ways, e.g., assigning each shared coefficient to the participant with the lowest identity index (assuming that the system operator has identity index 0). Notice that the mutual exclusiveness property above (i.e., $\tilde C_\Phi^i\cap\tilde C_\Phi^j=\emptyset$ for any $i\neq j$) is not possessed by the sets $C_\Phi^i$'s. This property can guarantee that each coefficient is only encrypted once in the secure computation algorithm (see line 3 of Algorithm \ref{algo: ITS DOP}). This is needed as the privacy of a coefficient could be compromised if it is encrypted multiple times (see the next subsection). For each $i\in\mathcal{V}\cup\{\rm SO\}$, let $m_i=|\tilde C_\Phi^i|$ and denote by $c_i=[c_{i\ell}]\in\mathbb{R}^{m_i}$ the vector of the elements of $\tilde C_\Phi^i$. Let $c =[c_i]_{i\in\mathcal{V}\cup\{\rm SO\}}\in\mathbb{R}^m$ with $m =\sum_{i\in\mathcal{V}\cup{\rm SO}}m_i$ and let $c_\mathcal{V}=[c_i]_{i\in\mathcal{V}}$.

For each $\Phi_{i\ell}$, we view it as a function of both $x$ and $c$. Denote by $\Phi_{i\ell}^s:\mathbb{R}^n\times\mathbb{R}^{m}\to\mathbb{R}$ the function with the same structure of $\Phi_{i\ell}$ but takes the coefficients $c$ also as variables and denote by $y=[y_i]\in\mathbb{R}^{m}$ the variables corresponding to $c$, where for each $i\in\mathcal{V}\cup{\rm SO}$, $y_i\in\mathbb{R}^{m_i}$ is the variable corresponding to $c_i$. The relation between $\Phi_{i\ell}$ and $\Phi_{i\ell}^s$ is then $\Phi_{i\ell}(x)=\Phi_{i\ell}^s(x,y)|_{y=c}$.

For each $\Phi_{i\ell}^s$, assume that it is written in the canonical form, i.e., sum of monomials, and denote by $\kappa_{i\ell}$ the number of monomials of $\Phi_{i\ell}^s$ and $Q_{i\ell}^v:\mathbb{R}^n\times\mathbb{R}^{m}\to\mathbb{R}$ the $v$-th monomial of $\Phi_{i\ell}^s$. Then, $\Phi_{i\ell}^s$ can be written as $\Phi_{i\ell}^s(x,y)=\sum_{v=1}^{\kappa_{i\ell}}Q_{i\ell}^v(x,y)$.

\subsection{An illustrative example\label{section private example}}

We first use an illustrative example to demonstrate the major challenges in the algorithm design. Consider the scenario of two agents and a system operator. Each agent $i\in\{1,2\}$ has a scalar state $x_i\in\mathbb{R}$. The joint functions of the two agents are $\Phi_1(x_1,x_2)=c_1x_1^2+c_2x_1x_2+c_3x_1$ and $\Phi_2(x_1,x_2)=c_1c_2x_2$, respectively, where $c_1,c_2,c_3\in\mathbb{R}$ are the private coefficients of agent 1 and $c_2\in\mathbb{R}$ is the private coefficient of agent 2.
%i.e., $C_\Phi^1=\{c_1,c_2,c_3\}$, $C_\Phi^2=\{c_2,c_4\}$ and $C_\Phi^{\rm SO}=\emptyset$. Assume $x_1=3.57$, $x_2=-6.42$, $c_1=1.25$, $c_2=2.36$, $c_3=-0.87$ and $c_4=5.34$. Since all the numbers have two decimal fraction digits, we can set $\sigma=2$. We aim to securely compute $\Phi_1(x_1,x_2)$ and $\Phi_2(x_1,x_2)$. For this example, one can check $\Phi_1(x_1,x_2)=$ and $\Phi_2(x_1,x_2)=$ (keeping two decimal fraction digits).

%The above example includes both addition and multiplication operations. Hence, we need to adopt a fully homomorphic encryption scheme. To the best of our knowledge, there does not exist a fully homomorphic encryption scheme for integers which is both efficiently implementable and semantically secure. Instead, there exist fully homomorphic encryption schemes which have weaker security levels but are efficiently implementable.  The SingleMod encryption \cite{cryptoeprint:2014:670} is an efficiently implementable (private key) fully homomorphic encryption scheme (for non-negative integers).
We next illustrate why existing homomorphic encryption schemes cannot be directly used to solve the above problem and clarify how the challenges are addressed.
%Existing homomorphic encryption schemes only work for binary or non-negative integers and hence cannot be directly applied to the above problem, where the coefficients and states are signed real numbers. The algorithm proposed in this section integrates the integer-real transformation mechanism \eqref{integer to real} and a redesign of the so-called SingleMod encryption \cite{cryptoeprint:2014:670}. We choose the SingleMod encryption as our designing prototype because it is an efficiently implementable (private key) fully homomorphic encryption scheme (for non-negative integers). We next illustrate the major challenges in applying the SingleMod encryption.
%As a standard requirement for a private key homomorphic encryption scheme, we assume that agent 1 and agent 2 have agreed on a positive integer key $w$ and keep it secret from the system operator.

(i) Transformation between real numbers and integers. Existing homomorphic encryption schemes only work for binary numbers or non-negative integers and hence cannot be directly applied to the above problem, where the coefficients $(c_1,c_2,c_3)$ and the states $(x_1,x_2)$ are all signed real numbers. We overcome this challenge by proposing the transformation mechanism \eqref{integer to real} between integers and real numbers. Roughly speaking, before encrypting a real data item, say $c_1$, with $\sigma$ decimal fraction digits, one first transforms it into an integer $10^\sigma c_1$. One can then apply a specific homomorphic encryption scheme, which will be discussed in the next paragraphs, to perform encryption, computation and decryption operations over the transformed integers. After that, the decrypted integer result is transformed back to a signed real number by the transformation mechanism \eqref{integer to real}.
%Lemma \ref{T function property} provides a theoretical guarantee for the transformation correctness. In particular, Lemma \ref{T function property} only holds when $w$ is larger than a lower bound. Accordingly, in our algorithm design, we need to set a lower bound for the key $w$ in order to guarantee the transformation correctness. Please refer to Assumption \ref{asm: key large enough} and Remark \ref{remark:large key}.

(ii) Sum operation over transformed integers. As mentioned in the last paragraph, before encrypting a real data item, say $c_1$, with $\sigma$ decimal fraction digits, one first transforms it into an integer $10^\sigma c_1$. This essentially scales the magnitude of $c_1$ by $10^\sigma$ times. Now let agent 1 form the monomials using the transformed integers $(10^\sigma x_1,10^\sigma c_1,10^\sigma c_2,10^\sigma c_3)$. Since $Q_1^1$ and $Q_1^2$ have degree 3, they are scaled by $10^{3\sigma}$ times, i.e., $Q_1^1(10^\sigma x,10^\sigma c)=10^\sigma c_1\cdot(10^\sigma x_1)^2=10^{3\sigma}c_1x_1^2$ and $Q_1^2(10^\sigma x,10^\sigma c)=10^\sigma c_2\cdot10^\sigma x_1\cdot10^\sigma x_2=10^{3\sigma}c_2x_1x_2$. Since $Q_1^3$ has degree 2, it is scaled by $10^{2\sigma}$ times, i.e., $Q_1^3(10^\sigma x,10^\sigma c)=10^\sigma c_3\cdot10^\sigma x_1=10^{2\sigma}c_3x_1$. Hence, with the transformed integers, the monomials $Q_1^1$, $Q_1^2$ and $Q_1^3$ are scaled by different times and thus the sum operation between them cannot be directly applied. We overcome this challenge by multiplying each monomial $Q_i^\ell$ formed with the transformed integers by a scaling term $10^{({\rm deg}(\Phi_i^s)-{\rm deg}(Q_i^\ell))\sigma}$. In particular, the scaled $Q_1^1$, $Q_1^2$ and $Q_1^3$ become $\hat Q_1^1=10^{(3-3)\sigma}10^{3\sigma}c_1x_1^2=10^{3\sigma}c_1x_1^2$, $\hat Q_1^2=10^{(3-3)\sigma}10^{3\sigma}c_2x_1x_2=10^{3\sigma}c_2x_1x_2$ and $\hat Q_1^3=10^{(3-2)\sigma}10^{2\sigma}c_3x_1=10^{3\sigma}c_3x_1$, respectively. Notice that the scaled monomials $(\hat Q_1^1,\hat Q_1^2,\hat Q_1^3)$ remain integers and simultaneously they are all scaled by $10^{3\sigma}$ times and hence the sum operation between them can be applied. The above method is generalized in line 6 of Algorithm \ref{algo: ITS DOP}.

(iii) Encryption of coefficients held by multiple agents. In the above example, $c_2$ is a private coefficient of both agent 1 and agent 2. A question is whether the security of $c_2$ could be compromised when it is encrypted by both agent 1 and agent 2. To answer this question, we first need to choose a homomorphic encryption scheme. The above example includes both addition and multiplication operations. Hence, we need to use a fully homomorphic encryption scheme. To the best of our knowledge, there does not exist a fully homomorphic encryption scheme for integers which is both efficiently implementable and semantically secure. Please refer to Section \ref{literature review on HE} for a summary of existing homomorphic encryption schemes. In this section, we choose the SingleMod encryption \cite{cryptoeprint:2014:670} as our designing prototype, because it is an \emph{efficiently implementable} (private key) fully homomorphic encryption scheme (for integers). On the other hand, the SingleMod encryption is not semantically secure. Instead, its security level adopts the standard notion of \emph{Plaintexts not efficiently solvable} (see Section \ref{security issues}). As mentioned in Section \ref{security issues}, many widely used encryption schemes adopt this notion of security, e.g., RSA, DES and AES. Since the SingleMod encryption is not semantically secure, the security of $c_2$ could be compromised if it is encrypted by both agent 1 and agent 2. For the SingleMod encryption, agent 1 and agent 2 agree on a large positive integer key $w$ and keep it secret from the system operator. To encrypt a non-negative integer $t<w$, one chooses an arbitrary positive integer $u_t$ and encrypts $t$ by $\hat t=u_tw+t$. Now assume that agent 1 and agent 2 encrypt $10^\sigma c_2$ (recall that $10^\sigma c_2$ is an integer) by $\hat y_2'=u_{y_2}'w+10^\sigma c_2$ and $\hat y_2''=u_{y_2}''w+10^\sigma c_2$, respectively, where $u_{y_2}'$ (resp. $u_{y_2}''$) is an arbitrary positive integer only known to agent 1 (resp. 2). By knowing $\hat y_2'$ and $\hat y_2''$, the system operator can compute $\hat y_2'-\hat y_2''=(u_{y_2}'-u_{y_2}'')w$. Notice that $u_{y_2}'$, $u_{y_2}''$, $\hat y_2'$, $\hat y_2''$ and $w$ are all integers. The system operator can then tell that $w$ must be a factor of $\hat y_2'-\hat y_2''$. In the worst case where $u_{y_2}'-u_{y_2}''$ and $w$ are both prime numbers, there are only two possible values for $w$. Assume $|10^\sigma c_2|\leq(w-1)/2$ (which is usually true as $w$ needs to be chosen very large for the concern of security). By Lemma \ref{T function property}, for each possible $w$, since the system operator knows the values of $\sigma$ and $\hat y_2'$, it can obtain a unique value of $c_2$ by computing $c_2=T_{\sigma,w}(10^\sigma c_2\mod w)=T_{\sigma,w}(\hat y_2'\mod w)$. By Definition \ref{def encryption semantic security}, this indicates that the SingleMod encryption is not semantically secure. By the above procedure, the system operator can obtain a good estimate of $c_2$.
%The above discussion indicates that the security of a private coefficient could be compromised if it is encrypted by multiple agents.
We overcome this challenge by constructing a partition of the coefficient set $C_\Phi$ as illustrated in Section \ref{section polynomial preliminaries} so that it is guaranteed that each private coefficient is only encrypted once. For this example, the partition could be either $\tilde C_\Phi^1=\{c_1,c_2,c_3\}$ and $\tilde C_\Phi^2=\emptyset$ in which agent 1 encrypts $c_2$, or $\tilde C_\Phi^1=\{c_1,c_3\}$ and $\tilde C_\Phi^2=\{c_2\}$ in which agent 2 encrypts $c_2$.
%Please refer to line 3 of Algorithm \ref{algo: ITS DOP}.

(iv) Causal attacks by the system operator. Assume that the system operator can use the data received up to time instant $k$ to infer $x(k)$. Again, due to that the SingleMod encryption is not semantically secure, the system operator could succeed if it fully knows the update rule of $x(k)$. We now assume that $c_1$ and $c_2$ are private coefficients of the system operator, $X_2=\mathbb{R}$, $\gamma(1)=1$, and these are known to the system operator. By \eqref{SCP of DC}, the system operator can derive $x_2(2)=(1-c_1c_2)x_2(1)$. Now agent 2 encrypts $x_2(1)$ and $x_2(2)$ by $\hat x_2(1)=u_{x_2}(1)w+10^\sigma x_2(1)$ and $\hat x_2(2)=u_{x_2}(2)w+10^\sigma x_2(2)$, respectively. By knowing $x_2(2)=(1-c_1c_2)x_2(1)$, the system operator can obtain $\hat x_2(2)-(1-c_1c_2)\hat x_2(1)=(u_{x_2}(2)-(1-c_1c_2)u_{x_2}(1))w$ and know that $w$ is a factor of $\hat x_2(2)-(1-c_1c_2)\hat x_2(1)$. Similar to the discussion in (iii), by knowing $\sigma$, $\hat x_2(1)$ and $\hat x_2(2)$, the system operator can then obtain a good estimate of $x_2(1)$ and $x_2(2)$. We overcome this challenge by assuming that the system operator can only launch temporarily independent attacks. Please refer to Assumption \ref{asm: one shot}. This assumption has been widely used in database privacy. Please refer to Remark \ref{remark one shot attack} for the detailed discussion of this assumption.

\subsection{Algorithm design and analysis\label{section private algorithm}}

The private key secure computation algorithm for \eqref{SCP of DC} is presented by Algorithm \ref{algo: ITS DOP}. This algorithm is based on the SingleMod encryption \cite{MVD-CG-SH-VV:2010}, \cite{cryptoeprint:2014:670}. As mentioned in item (iii) of Section \ref{section private example}, the SingleMod encryption is an efficiently implementable private key fully homomorphic encryption scheme for integers. The encryption privacy of our encryption scheme shares the same privacy notion with that of the SingleMod encryption, i.e., \emph{Plaintexts not efficiently solvable} (see Section \ref{security issues}). In particular, for our case, breaking the scheme is as hard as the approximate greatest common divisor (GCD) problem~\footnote{Approximate GCD: Given polynomially many integers randomly chosen close to multiples of a large integer $p$, i.e., in the form of $a_i=pq_i+r_i$, where $p$, $q_i$'s and $r_i$'s are all integers, find the ``common near divisor'' $p$.}, which is widely believed to be NP-hard \cite{MVD-CG-SH-VV:2010}, \cite{cryptoeprint:2014:670}.

\begin{algorithm}[htbp]
\caption{Private key secure computation algorithm}\label{algo: ITS DOP}
\small

\textbf{1.} \textbf{Initialization:} Each agent $i$ chooses any $x_i(0)\in X_i$;

\textbf{2.} \textbf{Key agreement:} All agents agree on a large odd positive integer $w$ and keep $w$ secret from SO;

\textbf{3.} \textbf{Coefficient encryption:} Each agent $i$ chooses any $u_{y_i}\in\mathbb{Z}_{>0}^{m_i}$ and sends to SO $\hat y_i$ computed as $\hat y_{i}=u_{y_i}w+10^{\sigma}c_i$; SO forms $\hat y=[\hat y_j]_{j\in\mathcal{V}\cup\{\rm SO\}}$ with $\hat y_{\rm SO}=10^{\sigma}c_{\rm SO}$;

\textbf{4.} \textbf{while} {$k\geq0$}

\textbf{5.} \quad \textbf{State encryption:} Each agent $i$ chooses any $u_{x_i}(k)\in\mathbb{Z}_{>0}^{n_i}$ and sends to SO $\hat x_i(k)$ computed as $\hat x_{i}(k)=u_{x_i}(k)w+10^{\sigma}x_i(k)$; SO forms $\hat x(k)=[\hat x_j(k)]_{j\in\mathcal{V}}$;

\textbf{6.} \quad \textbf{Computation over ciphertexts:} For each $i\in\mathcal{V}$, for each $\ell=1,\cdots,n_i$, SO sends $\bar \Phi_{i\ell}^s(k)$ to agent $i$ computed as:
\begin{align*}
\bar\Phi_{i\ell}^s(k)=&\sum\nolimits_{v=1}^{\kappa_{i\ell}}[10^{({\rm deg}(\Phi_{i\ell}^s)-{\rm deg}(Q_{i\ell}^v))\sigma}Q_{i\ell}^v(\hat x(k),\hat y)];
\end{align*}

\textbf{7.} \quad \textbf{Decryption:} For each $i\in\mathcal{V}$, for each $\ell=1,\cdots,n_i$, agent $i$ computes $\hat \Phi_{i\ell}^s(k) = T_{{\rm deg}(\Phi_{i\ell}^s)\sigma,w}(\bar\Phi_{i\ell}^s(k) \mod w)$;

\textbf{8.} \quad \textbf{Local update:} Each agent $i$ forms $\hat\Phi_{i}^s(k) =[\hat \Phi_{i\ell}^s(k)]$ and computes $x_i(k+1)=\mathbb{P}_{X_i}[x_i(k)-\gamma(k)\hat\Phi_{i}^s(k)]$;

\textbf{9.} \, Set $k\leftarrow k+1$;

\textbf{10.} \textbf{end while}
\end{algorithm}

The algorithm is informally stated next. At line 2, all the agents agree on a key $w$ and keep the value of $w$ unknown to the system operator. One case where such key agreement is possible is that the sub-communication graph between the agents is connected and all its communication links are secure~\footnote{A communication link between $i$ and $j$ is secure if only $i$ and $j$ can access the messages sent via this link while no others can access the messages.}.

For the sake of security, $w$ is chosen as a very large number, e.g., in the magnitude of $2^{2000}$ \cite{DG:2017}. On the other hand, to guarantee the correctness of decryption (line 7) which is a modulo operation over $w$, the value of $w$ cannot be too small. Roughly speaking, the value of $w$ must be larger than all possible plaintexts of computing results. This is captured by the following assumption.

\begin{assumption}
\label{asm: key large enough}
The key $w$ is chosen large enough such that, for any step $k$, it holds that $w\geq1+2\max_{i,\ell}10^{{\rm deg}(\Phi_{i\ell}^s)\sigma}|\Phi_{i\ell}(x(k))|$.\hfill\rule{1mm}{2mm}
\end{assumption}

\begin{remark}
\label{remark:large key}
A sufficient condition for Assumption \ref{asm: key large enough} is that $x$ lives in a compact set whose bound is known to the agents a priori. In the cases where this sufficient condition does not hold, since $w$ needs to be chosen very large for the sake of security, Assumption \ref{asm: key large enough} is usually automatically satisfied. In Section \ref{section case study}, we provide concrete examples for which this is true.\hfill\rule{1mm}{2mm}
\end{remark}

At line 3, each agent $i$ encrypts its coefficient $c_i$ via the SingleMod encryption using the private key $w$ and a random integer vector $u_{y_i}$. Recall that any involved real number has (at most) $\sigma$ decimal fraction digits. Hence, $10^\sigma c_i$ is an integer vector. Since $c_i$ is fixed throughout the computing process, it only has to be encrypted once.

At line 5, each agent $i$ encrypts its current state $x_i(k)$ via the SingleMod encryption using the key $w$ and a random integer vector $u_{x_i}(k)$. Notice that at each iteration, the encryption of $x_i(k)$ adopts a new randomness $u_{x_i}(k)$.

Line 6 is the computation step at which the system operator computes each function $\Phi_{i\ell}^s$ by the encrypted data $(\hat x(k),\hat y)$. For each $v=1,\cdots,\kappa_{i\ell}$, $Q_{i\ell}^v(\hat x(k),\hat y)$ is multiplied by $10^{({\rm deg}(\Phi_{i\ell}^s)-{\rm deg}(Q_{i\ell}^v))\sigma}$ in order to have each monomial scaled by the same times so that the sum operation can be performed. In particular, without the $u_{x_i}(k)w$ (resp. $u_{y_i}w$) part (which is eliminated by modulo operation in the decryption step), each $x_i(k)$ (resp. $c_i$) is scaled by $10^{\sigma}$ times in the integer transformation operation at line 5 (resp. 3). Each $Q_{i\ell}^v(\hat x(k),\hat y)$ is then scaled by $10^{{\rm deg}(Q_{i\ell}^v)\sigma}$ times and $10^{({\rm deg}(\Phi_{i\ell}^s)-{\rm deg}(Q_{i\ell}^v))\sigma}Q_{i\ell}^v(\hat x(k),\hat y)$ is scaled by $10^{{\rm deg}(\Phi_{i\ell}^s)\sigma}$ times. Thus, all the monomials of $\Phi_{i\ell}^s$ are scaled by the same times and can be summed up.

Line 7 is the decryption step. Each agent $i$ first performs modulo operation over the encrypted function value $\bar \Phi_{i\ell}^s(k)$ by the private key $w$ which eliminates all terms having $w$ as a factor. Then, agent $i$ transforms the resulted integer back into a signed real number by the $T$ function defined by \eqref{integer to real}.

Line 8 is the local update step. Each agent $i$ updates $x_i(k)$ by \eqref{SCP of DC} by using $\hat\Phi_i^s(k)$ as $\Phi_i(x(k))$.

To help with understanding, we provide a numerical example in Section \ref{section numerical example algorithm 1} to go through the steps of Algorithm \ref{algo: ITS DOP}.

\begin{assumption}
\label{asm: one shot}
The system operator can only perform temporarily independent attacks, i.e., at each step $k$, it uses the data received at step $k$ to infer $x(k)$, but does not use past data to collectively infer the sequence $x(0),\cdots,x(k)$.\hfill\rule{1mm}{2mm}
\end{assumption}

\begin{remark}
\label{remark one shot attack}
One case where Assumption \ref{asm: one shot} holds is that the system operator is not fully aware of the update rule of $x(k)$ and views the sequence $\{x(k)\}$ as a temporally independent time series \cite{ES-THHC-EGR-RC-DS:2011}. For example, the system operator could be an untrustworthy third party who does not know the underlying update rule \eqref{SCP of DC} or even does not know what problem the agents aim to solve \cite{SR-MX-SKD:2012}. In this scenario, the system operator only receives a sequence of (encrypted) aggregate results from an unknown system. This scenario has been widely considered in database privacy and many works in the field are based on the temporal independence assumption. For example, the work \cite{RB-AB-VG-SL-AT:2011}, which studied noiseless database privacy, explicitly made the assumption that the entries in the database are uncorrelated and justified the assumption by that, without knowing a system, observing aggregate information released from the system may provide little knowledge on the correlation of the entries. Moreover, as pointed out by \cite{CD-MN-TP-GNR:2010} and Chapter 14 of \cite{TZ-GL-WZ-PSY:2017}, most existing differential privacy works assume that the dataset consists of independent records, despite the fact that records in real world applications are often correlated.\hfill\rule{1mm}{2mm}
\end{remark}

\begin{theorem}
Suppose that Assumptions \ref{asm: star topology}, \ref{asm: polynomial}, \ref{asm: key large enough} and \ref{asm: one shot} hold. By Algorithm \ref{algo: ITS DOP}, the following claims hold:

1) Correctness: $\hat \Phi_i^s(k)=\Phi_i(x(k))$ for any $k$ and $i\in\mathcal{V}$.

2) Security: Between the agents, Algorithm \ref{algo: ITS DOP} securely computes the sequence of $\{\Phi_1(x(k)),\cdots,\Phi_N(x(k))\}$; for the system operator, at each step $k$, Algorithm \ref{algo: ITS DOP} is as hard as the approximate GCD problem for the system operator to solve $x(k)$ and $c_{\mathcal{V}}$.\hfill\rule{1mm}{2mm}
\label{theorem: GNEP}
\end{theorem}

The proof of Theorem \ref{theorem: GNEP} is given in Section \ref{proof of GNEP theorem}. In the proof, the privacy notion of \emph{plaintexts not efficiently solvable} (Section \ref{security issues}) and Definition \ref{secure compute def} are used. In particular, the correctness is derived by exploring the homomorphic property of the proposed private key encryption scheme and the property of the proposed real-integer transformation scheme (Lemma \ref{T function property}). The secure computation result between the agents is obtained by using the simulation paradigm which is a standard technique for proving SMC \cite{RC-ID-JBN:2015}, \cite{YL-BP:09}. The security level against the system operator is based on the fact that solving $x(k)$ and $c_{\mathcal{V}}$ is exactly the approximate GCD problem.

\begin{remark}
\label{remark not semantic security}
The encryption scheme of Algorithm \ref{algo: ITS DOP} leverages the SingleMod encryption. As pointed out in \cite{cryptoeprint:2014:670}, the SingleMod encryption is not semantically secure. Instead, the encryption privacy of the SingleMod encryption is claimed in the sense of Plaintexts not efficiently solvable. On the one hand, this privacy notion is weaker than semantic security \cite{XY-RP-EB:2014}. On the other hand, the weaker security level renders that Algorithm \ref{algo: ITS DOP} is fully homomorphic and efficiently implementable (see Section \ref{section simulation 1} for efficiency). As mentioned in Section \ref{security issues}, this sense of privacy claim has been adopted in many widely used encryption schemes, e.g., RSA, DES and AES. In the next section, we assume that the joint functions are affine and provide a partially homomorphic encryption scheme which possesses semantic security.
\hfill\rule{1mm}{2mm}
\end{remark}

\section{Public key secure computation algorithm\label{public key section}}

Algorithm \ref{algo: ITS DOP} proposed in the last section has several limitations: key distribution problem exists as the agents have to agree on a private key; its security level against the system operator is not semantic security, and as a consequence, it can only resist temporarily independent attacks against the system operator. In this section, we consider the special case where $\Phi_{i\ell}$'s are affine functions and propose a public key secure computation algorithm to address these limitations. To be more specific, the class of computation problems considered in this section is specified by the following assumption.

\begin{assumption}
For each $i\in\mathcal{V}$ and each $\ell\in\{1,\cdots,n_i\}$, the function $\Phi_{i\ell}$ is affine in $x$ and the coefficients of $\Phi_{i\ell}$ are all known to the system operator.\hfill\rule{1mm}{2mm}
\label{asm: special case of DOP}
\end{assumption}

\begin{remark}
Assumption \ref{asm: special case of DOP} indicates that each agent $i$'s coefficients are known to the system operator but may not be known to other agents.

The reason that Assumption \ref{asm: special case of DOP} is needed for this section is that the Paillier encryption scheme adopted in this section is only additively homomorphic but not multiplicatively homomorphic. Thus, we need the desired functions $\Phi_{i\ell}$'s to be affine in $x$. Moreover, if the function to be computed is a weighted sum of $x$, then the weights (transformed into integers) must be known by the entity who performs the computation (the system operator in our algorithm) so that the multiplication of a weight and a variable can be carried out by summing up the variable with itself for certain times (see (3-ii) of Theorem \ref{theorem: Paillier}). A similar treatment as Assumption \ref{asm: special case of DOP} was adopted in \cite{NF-PP:2016} in the context of encrypted average consensus.

To the best of our knowledge, all computationally efficient public key homomorphic encryption schemes are partially homomorphic, e.g., the RSA scheme \cite{RLR-AS-LA:1978} and the ElGamal scheme \cite{TE:1985} are multiplicatively homomorphic, and the Goldwasser--Micali scheme \cite{SG-SM:1982} and the Paillier scheme \cite{PP:1999} are additively homomorphic. The first fully public key homomorphic encryption scheme was developed by Gentry in his seminal work \cite{CG-2009}. However, due to that the Gentry fully homomorphic encryption scheme uses lattice and bootstrapping, its implementation is rather complicated and time-consuming, which significantly limits its applications in practice.

As mentioned in Section \ref{section SCP formulation}, $\Phi_i$ is the first-order gradient of certain functions with respect to $x_i$. A sufficient and necessary condition for the affinity assumption is that the associated component functions are affine or quadratic in $x_i$. Hence, problem \eqref{SCP of DC} satisfying Assumption \ref{asm: special case of DOP} covers a large class of problems, e.g., linear and quadratic programs \cite{Bertsekas.Tsitsiklis:97}, quadratic convex games \cite{MZ-EF:Auto14} and affine variational inequalities \cite{FF-JSP:03}.\hfill\rule{1mm}{2mm}
\end{remark}

By Assumption \ref{asm: special case of DOP}, each function $\Phi_{i\ell}$ is a weighted sum of $x$ for which the weights are known to the system operator. For convenience of notation, we write each $\Phi_{i\ell}$ as:
\begin{align}
\label{affine function for Paillier}
\Phi_{i\ell}(x)=\sum\nolimits_{j=1}^N\sum\nolimits_{v=1}^{n_j}A_{i\ell}^{jv}x_{jv}+B_{i\ell}
\end{align}
where $A_{i\ell}^{jv}\in\mathbb{R}$ and $B_{i\ell}\in\mathbb{R}$ are coefficients known to the system operator. For each $j\in\mathcal{V}$, let $A_{i\ell}^j=[A_{i\ell}^{j1},\cdots,A_{i\ell}^{jn_j}]$ and $A_i^j=[A_{i\ell}^{j}]_{\ell\in\{1,\cdots,n_i\}}$. For each $i\in\mathcal{V}$, let $A_{i\ell}=[A_{i\ell}^1,\cdots,A_{i\ell}^N]$ for each $\ell=1,\cdots,n_i$, $A_i = [A_i^1,\cdots,A_i^N]$ and $B_i = [B_{i1},\cdots,B_{in_i}]^T$. Let $A = [A_i]_{i\in\mathcal{V}}$ and $B = [B_i]_{i\in\mathcal{V}}$. By these notations, we have $\Phi_{i\ell}(x)=\sum\nolimits_{j=1}^NA_{i\ell}^jx_j+B_{i\ell}=A_{i\ell}x+B_{i\ell}$ and $\Phi_i(x)=\sum\nolimits_{j=1}^NA_{i}^jx_j+B_i=A_ix+B_i$.

In this section, we use the Paillier encryption scheme as our design prototype. This is because the Paillier encryption scheme is an efficiently implementable (public key) additively homomorphic encryption scheme (for integers). Again, we use the proposed transformation mechanism \eqref{integer to real} to overcome the challenge of signed real numbers.
%According to the requirement of Lemma \ref{T function property}, we set a lower bound for the Paillier keys in order to guarantee the transformation correctness. Please refer to Assumption \ref{asm: public key large enough}.
The other challenges illustrated in Section \ref{section private example} do not exist for this case. Since the monomials $A_{i\ell}^{jv}x_{jv}$'s have the same degree, challenge (ii) in Section \ref{section private example} does not exist. Since we assume that the coefficients $A_{i\ell}^{jv}$'s and $B_{i\ell}$'s are all known to the system operator, the agents do not have to encrypt any coefficient and hence challenge (iii) in Section \ref{section private example} does not exist. Since the Paillier encryption scheme is semantically secure, challenge (iv) in Section \ref{section private example} does not exist. Informally speaking, this is because even if the system operator fully knows the relation between some $x_i(k)$ and $x_i(k+1)$, by Definition \ref{def encryption semantic security}, the encryptions of $x_i(k)$ and $x_i(k+1)$ provide zero knowledge to the system operator.

\subsection{Algorithm design and analysis}

The algorithm developed in this section is based on a public key partially homomorphic encryption scheme, namely, the Paillier encryption scheme \cite{PP:1999}. The preliminaries for the Paillier encryption scheme are provided in Section \ref{Paillier priliminaries}.

\begin{algorithm}[htbp]
\caption{Public key secure computation algorithm}\label{algo: semantically secure DOP}
\small

\textbf{1.} \textbf{Initialization:} Each agent $i$ chooses any $x_i(0)\in X_i$;

\textbf{2.} \textbf{Key generation:} Each agent $i$ generates Paillier keys $(\alpha_i,\beta_i,\nu_i,\pi_i)$, publicizes $(\alpha_i,\beta_i)$ while keeps $(\nu_i,\pi_i)$ private;

\textbf{3.} \textbf{while} {$k\geq0$}

\textbf{4.} \quad \textbf{State encryption:} For each $i\in\mathcal{V}$ and $\ell\in\{1,\cdots,n_i\}$, for each $j\in\mathcal{V}$, agent $i$ selects any $r_{i\ell}^j(k)\in\mathbb{Z}_{\alpha_j}^*$ and sends to SO $\hat x_{i\ell}^j(k)=\beta_j^{10^{\sigma}x_{i\ell}(k)\mod \alpha_j}\cdot r_{i\ell}^j(k)^{\alpha_j}\mod\alpha_j^2$;

\textbf{5.} \quad \textbf{Computation over ciphertexts:} For each $i\in\mathcal{V}$ and $\ell\in\{1,\cdots,n_i\}$, SO sends $\bar \Phi_{i\ell}^s(k)$ to agent $i$ computed as:
\begin{align*}
\bar \Phi_{i\ell}^s(k)=&(\beta_i^{10^{2\sigma}B_{i\ell} \!\!\!\!\!\mod \!\alpha_i}\prod\limits_{j=1}^N\prod\limits_{v=1}^{n_j}\hat x_{jv}^i(k)^{10^{\sigma}A_{i\ell}^{jv} \!\!\!\!\!\mod \!\alpha_i}) \!\!\!\!\!\mod \!\alpha_i^2;
\end{align*}

\textbf{6.} \quad \textbf{Decryption:} For each $i\in\mathcal{V}$ and $\ell\in\{1,\cdots,n_i\}$, agent $i$ computes $\hat \Phi_{i\ell}^s(k) = T_{2\sigma,\alpha_i}(\frac{(\bar \Phi_{i\ell}^s(k)^{\nu_i} \mod \alpha_i^2)-1}{\alpha_i}\cdot\pi_i \mod \alpha_i)$;

\textbf{7.} \quad \textbf{Local update:} Each agent $i$ forms $\hat\Phi_{i}^s(k) = [\hat \Phi_{i\ell}^s(k)]$ and computes $x_i(k+1)=\mathbb{P}_{X_i}[x_i(k)-\gamma(k)\hat\Phi_{i}^s(k)]$;

\textbf{8.} \, Set $k\leftarrow k+1$;

\textbf{9.} \textbf{end while}
\end{algorithm}

The public key secure computation algorithm for \eqref{SCP of DC} satisfying Assumption \ref{asm: special case of DOP} is presented by Algorithm \ref{algo: semantically secure DOP}.

At line 2, each agent $i$ generates its Paillier keys $(\alpha_i,\beta_i,\nu_i,\pi_i)$, where $(\alpha_i,\beta_i)$ are public keys known to all participants while $(\nu_i,\pi_i)$ are private keys only known to agent $i$. For the sake of security, the keys are chosen very large, e.g., in the magnitude of $2^{2000}$ \cite{DG:2017}. To guarantee decryption correctness, we need the following Assumption \ref{asm: public key large enough}. Remark \ref{remark:large key} applies to this assumption.

\begin{assumption}
\label{asm: public key large enough}
For each agent $i$, its public key $\alpha_i$ is chosen large enough such that, for any step $k$, it holds that $\alpha_i\geq1+2\times10^{2\sigma}\|A_ix(k)+B_i\|_{\max}$.\hfill\rule{1mm}{2mm}
\end{assumption}

Line 4 is the encryption step. Each agent $i$ encrypts $x_i(k)$ by the Paillier encryption operation $N$ times such that the $j$-th encryption uses agent $j$'s public keys $(\alpha_j,\beta_j)$ and the corresponding encrypted data $\hat x_i^j(k)$ is used for the computation of agent $j$'s desired functions $\Phi_{j}(x(k))$. By receiving $\hat x_{i\ell}^j(k)$ and only knowing agent $j$'s public keys $(\alpha_j,\beta_j)$, under the decisional composite
residuosity assumption (DCRA) \footnote{The statement of the DCRA is given in Section \ref{Paillier priliminaries}.}, it is computationally intractable for the system operator to infer $x_{i\ell}(k)$.

Line 5 is the computation step. The computation performed by the system operator is based on the additively homomorphic property of the Paillier encryption scheme, i.e., roughly speaking, the multiplication of ciphertexts provides the encryption of the sum of the corresponding plaintexts (see Section \ref{Paillier priliminaries}).

Line 6 is the decryption step. Agent $i$ decrypts each $\bar \Phi_{i\ell}^s(k)$ by the Paillier decryption operation and transforms the decrypted non-negative integer into a signed real number by the $T$ function defined by \eqref{integer to real}.

Line 7 is the local update step. Each agent $i$ updates $x_i(k)$ by \eqref{SCP of DC} by using $\hat\Phi_i^s(k)$ as $\Phi_i(x(k))$.

To help with understanding, we provide a numerical example in Section \ref{section numerical example algorithm 2} to go through the steps of Algorithm \ref{algo: semantically secure DOP}.

\begin{theorem}
Suppose that Assumptions \ref{asm: star topology}, \ref{asm: special case of DOP}, \ref{asm: public key large enough} and the standard cryptographic assumption DCRA hold. By Algorithm \ref{algo: semantically secure DOP}, the following claims hold:

1) Correctness: $\hat \Phi_i^s(k)=\Phi_i(x(k))$ for any $k$ and $i\in\mathcal{V}$.

2) Security: Algorithm \ref{algo: semantically secure DOP} securely computes the sequence of $\{\Phi_1(x(k)),\cdots,\Phi_N(x(k))\}$ between the agents and is semantically secure against the system operator.\hfill\rule{1mm}{2mm}
\label{theorem: DOP special case}
\end{theorem}

The proof of Theorem \ref{theorem: DOP special case} is given in Section \ref{proof of DOP special case theorem}. In the proof, Definitions \ref{def encryption semantic security} and \ref{secure compute def} are used. In particular, the correctness is derived by the homomorphic properties of the Paillier encryption scheme (part (3) of Theorem \ref{theorem: Paillier}) and the property of the proposed real-integer transformation scheme (Lemma \ref{T function property}). The secure computation result between agents is obtained by using the simulation paradigm. The semantic security against the system operator follows the semantic security of the Paillier encryption scheme (combination of part (2) of Theorem \ref{theorem: Paillier} and Theorem 5.2.10 of \cite{OG:2004}).

\section{Privacy analysis on input-output inference\label{observability section}}

In the last two sections, we provide two algorithms to achieve SMC. However, as pointed out in Section \ref{section privacy issues}, it is possible that the adversary could infer the private inputs of the securely computed functions purely from its own inputs and the function outputs. In this section, we study the second problem raised in Section \ref{section privacy issues}, that is, whether the private inputs of a function can be uniquely determined from its outputs.

Recall that $\Phi_{i}^s$ denotes the function with the same structure of $\Phi_i$ but takes the coefficients $c$ also as variables (see Section \ref{section polynomial preliminaries}). Denote by $(\{\bar x(k)\},\bar X,\bar c)$ the true values of $(\{x(k)\},X,c)$. For each $i\in\mathcal{V}$, agent $i$ aims to find feasible $(\{\hat x_{-i}(k)\},\hat X_{-i},\hat c_{-i})$ that satisfy its observations: $\Phi_i^s(\bar x_i(k),\hat x_{-i}(k),\bar c_i,\hat c_{-i})=\Phi_i(\bar x(k))$ for all $k\in\mathbb{N}$, and $\hat x_j(k+1)=\mathbb{P}_{\hat X_j}[\hat x_j(k)-\gamma(k)\Phi_j^s(\bar x_i(k),\hat x_{-i}(k),\bar c_i,\hat c_{-i})]$ for all $j\in\mathcal{V}\backslash\{i\}$ and all $k\in\mathbb{N}$. By Definition \ref{secure from output}, the sequence of functions $\{\Phi(x(k))\}$ resists input-output inference with unbounded uncertainty if, for any agent $i\in\mathcal{V}$, each element of the feasible set $\{\{\hat x_{-i}(k)\},\hat X_{-i},\hat c_{-i}\}$ has unbounded uncertainty.

In this section, we study the scenario where $\Phi_{i\ell}$'s are a class of quadratic functions specified by the following assumption.

\begin{assumption}
\label{asm: IOI problem}
For each $i\in\mathcal{V}$ and each $\ell=1,\cdots,n_i$, $\Phi_{i\ell}(x)=x^TH_{i\ell}x+A_{i\ell}x+B_{i\ell}$, where $H_{i\ell}\in\mathbb{R}^{n\times n}$ and $A_{i\ell}\in\mathbb{R}^{1\times n}$ are public constant matrix and vector, respectively, that are known to all the agents, while $B_{i\ell}\in\mathbb{R}$ is a private constant scalar only known to a subset of agents.\hfill\rule{1mm}{2mm}
\end{assumption}

\subsection{An illustrative example\label{section IOI example}}

We first use an illustrative example to show the issue of input-output inference. Consider the case of three agents in which agent 1 is adversary and agent 2 and agent 3 are benign. Each agent $i\in\{1,2,3\}$ has a scalar state $x_i\in\mathbb{R}$. The joint functions of the three agents are $\Phi_1(x_1,x_2,x_3)=-x_2-x_3$, $\Phi_2(x_1,x_2,x_3)=-2x_3$ and $\Phi_3(x_1,x_2,x_3)=-x_1$, respectively. For ease of presentation, assume $X_i=\mathbb{R}$ for all $i\in\{1,2,3\}$ and $\gamma(k)=1$ for all $k\in\mathbb{N}$, and these are known to all the three agents. By \eqref{SCP of DC}, we then have $x_1(k+1)=x_1(k)+x_2(k)+x_3(k)$, $x_2(k+1)=x_2(k)+2x_3(k)$ and $x_3(k+1)=x_1(k)+x_3(k)$. We next show that by knowing $x_1(k)$, $x_1(k+1)$ and $x_1(k+2)$, agent 1 can uniquely determine $x_2(k)$ and $x_3(k)$. Agent 1 knows $x_1(k+1)=x_1(k)+x_2(k)+x_3(k)$ and $x_1(k+2)=x_1(k+1)+x_2(k+1)+x_3(k+1)$. By plugging the relations $x_2(k+1)=x_2(k)+2x_3(k)$ and $x_3(k+1)=x_1(k)+x_3(k)$ into $x_1(k+2)=x_1(k+1)+x_2(k+1)+x_3(k+1)$, agent 1 derives $x_2(k)+3x_3(k)=x_1(k+2)-x_1(k+1)-x_1(k)$. Together with the equation $x_2(k)+x_3(k)=x_1(k+1)-x_1(k)$, agent 1 can uniquely derive $x_2(k)=-\frac{1}{2}(x_1(k+2)-4x_1(k+1)+2x_1(k))$ and $x_3(k)=\frac{1}{2}(x_1(k+2)-2x_1(k+1))$.

The above issue arises because the process defined by \eqref{SCP of DC} is iterative so that an adversarial agent can use successive observations to infer the private inputs of the other agents. Existing approaches in the privacy literature, e.g., \cite{MRC-ACM-FBS:2009}, \cite{PM-SM-MH-MS:2011}, \cite{PM-MH-JK-MS:2012}, are only applicable to problems with constant private inputs, but inapplicable to the problem concerned in our paper, in which the private inputs are time series generated by a dynamic system. Hence, the iterative nature of the process necessitates new analysis approaches.

We next modify the above example as $x_1(k+1)=x_1(k)+x_2(k)+x_3(k)$, $x_2(k+1)=x_1(k)+x_3(k)$ and $x_3(k+1)=x_1(k)+x_2(k)$. Following the above procedure, agent 1 can derive $x_1(k+1)=x_1(k)+x_2(k)+x_3(k)$ and $x_1(k+2)=x_1(k+1)+2x_1(k)+x_2(k)+x_3(k)$. Notice that these two equations degenerate to one in terms of the unknowns $x_2(k)$ and $x_3(k)$. Hence agent 1 cannot determine the two unknowns $x_2(k)$ and $x_3(k)$ from only one equation. Actually, one can check by mathematical induction that, no matter how many observations agent 1 uses, the equations it derives all degenerate to the one above and hence it cannot determine $x_2(k)$ and $x_3(k)$.

The two examples above indicate that the issue of input-output inference depends on the structural properties of the joint functions. We will identify a sufficient condition on the structural properties of the null space of the weight matrices of the joint functions under which the affine functions resist input-output inference. The formal analysis is presented in the next subsection.

\subsection{Analysis\label{section IOI analysis}}

The quadratic term is decomposed as $x^TH_{i\ell}x=\sum_{u,v=1}^Nx_u^TH_{i\ell}^{uv}x_v$, where each $H_{i\ell}^{uv}$ is a block sub-matrix of $H_{i\ell}$. Denote $H_{i\ell}^v=[H_{i\ell}^{uv}+(H_{i\ell}^{vu})^T]_{u\in\mathcal{V}}$ for each $v\in\mathcal{V}$. For each $i\in\mathcal{V}$, denote $H_{j\ell}^{-i}={\rm diag}\{H_{j\ell}^1,\cdots,H_{j\ell}^{i-1},H_{j\ell}^{i+1},\cdots,H_{j\ell}^N\}$ for each $j\in\mathcal{V}$ and $\ell=1,\cdots,n_j$. Denote $H^{-i}=[H_{j\ell}^{-i}]_{j\in\mathcal{V},\ell\in\{1,\cdots,n_j\}}$, i.e., $H^{-i}$ is obtained by stacking the quadratic weight matrices of $x_{-i}$ in all $\Phi_{j\ell}$'s. For each $i\in\mathcal{V}$, let $\Omega_i=\{(j,\ell):{\rm agent\;}i\;{\rm knows\;}B_{j\ell}\}$, $\Omega_i'=\{(j,\ell):{\rm agent\;}i\;{\rm does\;not\;know\;}B_{j\ell}\}$ and $A_{j\ell}^{-i}=[A_{j\ell}^1,\cdots,A_{j\ell}^{i-1},A_{j\ell}^{i+1},\cdots,A_{j\ell}^{N}]$ for each $j\in\mathcal{V}$ and $\ell=1,\cdots,n_j$. Then $A^{-i}=[A_{j\ell}^{-i}]_{(j,\ell)\in\Omega_i}$ is the matrix obtained by stacking the weight matrices of $x_{-i}$ in all those $\Phi_{j\ell}$'s such that the constant term $B_{j\ell}$'s are known to agent $i$. For example, let $N=4$, $n_j=3$ for all $j=1,\cdots,4$, $\Omega_1=\{(1,3),(2,1),(3,2),(4,2)\}$ and $\Omega_1'=\{(1,1),(1,2),(2,2),(2,3),(3,1),(3,3),(4,1),(4,3)\}$. In this case, $H_{j\ell}^{-1}={\rm diag}\{H_{j\ell}^2,H_{j\ell}^3,H_{j\ell}^4\}$, $A_{j\ell}^{-1}=[A_{j\ell}^2,A_{j\ell}^3,A_{j\ell}^4]$ for all $(j,\ell)$, $H^{-1}=[(H_{11}^{-1})^T,(H_{12}^{-1})^T,$ $(H_{13}^{-1})^T,\cdots,(H_{41}^{-1})^T,(H_{42}^{-1})^T,(H_{43}^{-1})^T]^T$ and $A^{-1}=[(A_{13}^{-1})^T,(A_{21}^{-1})^T,(A_{32}^{-1})^T,$ $(A_{42}^{-1})^T]^T$.

\begin{assumption}
For each $i\in\mathcal{V}$, there exists a null vector $o_i$ of $[(H^{-i})^T,(A^{-i})^T]^T$ such that each entry of $o_i$ is nonzero.\hfill\rule{1mm}{2mm}
\label{asm: input output inference}
\end{assumption}

\begin{remark}
If $\Omega_i = \emptyset$ for some $i\in\mathcal{V}$, then $A^{-i}$ can be removed from Assumption \ref{asm: input output inference}. For the affine case of Section \ref{public key section} where $\Phi_{i\ell}(x)=A_{i\ell}x+B_{i\ell}$, $H^{-i}$ can be removed from Assumption \ref{asm: input output inference}. Thus, the result of this section includes affine functions as a special case.\hfill\rule{1mm}{2mm}
\end{remark}

\begin{theorem}
Suppose Assumption \ref{asm: input output inference} holds. The sequence of functions $\{\Phi(x(0)),\cdots,\Phi(x(K))\}$ specified by Assumption \ref{asm: IOI problem} resists input-output inference with unbounded uncertainty for any $K\in\mathbb{N}$.\hfill\rule{1mm}{2mm}
\label{IO resist linear case I}
\end{theorem}

The proof of Theorem \ref{IO resist linear case I} is given in Section \ref{proof of IO resist linear case I}. In the proof, Definition \ref{secure from output} is used. Here we provide some intuitions behind the theorem. We fix an adversary agent $i$ and aim to construct $(\{x_{-i}(k)\},X_{-i},[B_{j\ell}]_{(j,\ell)\in\Omega_i'})$ that satisfy the sequence of its observations. For each $j\in\mathcal{V}$, denote the true $(X_j,B_{j},\{x_j(k)\})$ by $(\bar X_j,\bar B_j,\{\bar x_j(k)\})$. A feasible construction is: $\hat x_j(k)=\bar x_j(k)+\delta_j$ for each $j\in\mathcal{V}$ and each step $k$, where $\delta_i=0$, and for each $j\neq i$, $\delta_j\in\mathbb{R}^{n_j}$ is an arbitrary constant vector such that each entry of each $\delta_j$ is nonzero and $[(H^{-i})^T,(A^{-i})^T]^T\delta_{-i}=0$, with $\delta_{-i}=[\delta_v]_{v\neq i}$; $\hat X_j = \{x_j+\delta_j:x_j\in \bar X_j\}$ for each $j\neq i$; $\hat B_{j\ell}=\bar B_{j\ell}-A_{j\ell}\delta$ for each $(j,\ell)\in\Omega_i'$ with $\delta=[\delta_v]_{v\in\mathcal{V}}$. Assumption \ref{asm: input output inference} guarantees the existence of the above set of $\delta_j$'s. Informally speaking, for each $(j,\ell)\in\Omega_i$, the overall perturbation of $\delta_v$'s is absorbed by the relation $[(H^{-i})^T,(A^{-i})^T]^T\delta_{-i}=0$ and the construction of $\hat X_{j\ell}$; for each $(j,\ell)\in\Omega_i'$, the perturbation is absorbed by the constructions of $\hat B_{j\ell}$ and $\hat X_{j\ell}$ and the relation $H^{-i}\delta_{-i}=0$. It would be easy to check that, for any nonzero real number $r$, since $r\delta$ also satisfies the condition given in Assumption \ref{asm: input output inference}, i.e., $[(H^{-i})^T,(A^{-i})^T]^Tr\delta_{-i}=0$, the set of $r\delta_j$'s is also feasible with the same $\hat X_j$'s and $\hat B_j$'s. Since each entry of each $\delta_j$ is nonzero, each entry of each $r\delta_j$ is also nonzero and could be arbitrarily large by choosing $r$ arbitrarily large. By this, we see that the above construction has an infinite number of possibilities and the set of such constructions is unbounded. See Section \ref{proof of IO resist linear case I} for the formal proof of the above reasoning.

\section{Case study\label{section case study}}

In this section, we validate the efficacy of Algorithm \ref{algo: ITS DOP} and Algorithm \ref{algo: semantically secure DOP} by a demand response problem and an optimal power flow problem, respectively.

The simulation environment for both two case studies is as follows. On the hardware side, the simulation is performed on a Dell OptiPlex 9020 desktop computer with Intel(R) Core(TM) i7-4790 CPU at 3.60 GHz. On the software side, the simulation is performed on MATLAB R2016a. The involved operations over big integers, including key generation, encryption, computation over encrypted data and decryption, are all performed over the type of java.math.BigInteger variables under the integrated Java 1.7.$0\_60-$b19 with Oracle Corporation Java HotSpot(TM) 64-Bit Server VM mixed mode.

\subsection{Algorithm \ref{algo: ITS DOP}\label{section simulation 1}}

We simulate Algorithm \ref{algo: ITS DOP} by a demand response problem.

\subsubsection{Demand response problem}

\emph{$\bullet$ Problem formulation.} Consider a power network which is modeled as an interconnected graph $\mathcal{G}_p=\{\mathcal{V}_p,\mathcal{E}_p\}$, where each node $i\in\mathcal{V}_p$ represents a bus and each link in $\mathcal{E}_p$ represents a line. Each bus is connected to either a power supply or load and each load is associated with an agent. The set of load buses is denoted by $\mathcal{V}_l\subseteq\mathcal{V}_p$ and the set of supply buses is denoted by $\mathcal{V}_s\subseteq\mathcal{V}_p$. In a demand response problem, the agents could be households/customers and the system operator could be a utility company that sells electricity to the customers \cite{NL-LC-SHL:2011}.

\begin{table}[!t]
\renewcommand{\arraystretch}{1}
\caption{Parameters/variables of the demand response problem}
\label{DR table}
\centering
\begin{tabular}{|c|c|}
\hline
$c$ & disutility cost parameters\\
\hline
$R$ & reduced load\\
\hline
$S$ & maximum available power supply\\
\hline
$L$ & power load\\
\hline
$f^{\rm max}$ & maximum line capacity\\
\hline
$H_s$ & generation shift factor matrix\\
\hline
$H_l$ & load shift factor matrix\\
\hline
\end{tabular}
\end{table}

The physical meanings of the parameters and variables of the demand response problem are listed in Table \ref{DR table}. Denote by $S_i\in\mathbb{R}_{\geq0}$ the maximum available power supply at bus $i\in\mathcal{V}_s$ and by $L_i\in\mathbb{R}_{\geq0}$ the intended power load at bus $i\in\mathcal{V}_l$. Let $S=[S_i]_{i\in\mathcal{V}_s}$ and $L=[L_i]_{i\in\mathcal{V}_l}$. If $\textbf{1}_{|\mathcal{V}_s|}^TS\geq\textbf{1}_{|\mathcal{V}_l|}^TL$, then all the intended loads can be satisfied. Otherwise, some customers have to reduce their loads. For each $i\in\mathcal{V}_l$, denote by $R_i\in[0,L_i]$ the reduced load of customer $i$ and let $R\buildrel \Delta \over =[R_i]_{i\in\mathcal{V}_l}$. Each customer $i$ aims to solve the the following optimization problem:
\begin{align}
\label{demand response}
&\!\!\!\!\!\!\!\!\!\!\!\mathop {\min }\limits_{R_i \in {[0,L_i]}} c_iR_i-\Lambda _i(L_i-R_i)+\Upsilon (\textbf{1}_{|\mathcal{V}_l|}^T(L-R))(L_i-R_i)\nonumber\\
{\rm{s.t.}}\,\,&\textbf{1}_{|\mathcal{V}_l|}^T(L-R)\leq\textbf{1}_{|\mathcal{V}_s|}^TS,\;H_sS-H_l(L-R)\leq f^{\rm max},\nonumber\\
&-H_sS+H_l(L-R)\leq f^{\rm max}
\end{align}
where $c_iR_i$ represents the disutility induced by customer $i$'s load reduction $R_i$ with $c_i>0$; $\Lambda_i:\mathbb{R}_{\geq0}\to\mathbb{R}$ is a scalar function and $\Lambda_i(L_i-R_i)$ is the benefit produced by customer $i$'s actual load $L_i-R_i$; $\Upsilon:\mathbb{R}_{\geq0}\to\mathbb{R}_{\geq0}$ and $\Upsilon(\textbf{1}_{|\mathcal{V}_l|}^T(L-R))$ stands for the charged price given the total actual load $\textbf{1}_{|\mathcal{V}_l|}^T(L-R)$; $H_s\in[-1,1]^{|\mathcal{E}_p|\times|\mathcal{V}_s|}$ (resp. $H_l\in[-1,1]^{|\mathcal{E}_p|\times|\mathcal{V}_l|}$) is the generation (resp. load) shift factor matrix such that the $(j,\ell)$-th entry represents the power distributed on line $j$ when $1MW$ is injected into (resp. withdrawn from) bus $\ell$; $f^{\rm max}=[f_j^{\rm max}]_{j\in\mathcal{E}_p}$ with $f_j^{\rm max}\in\mathbb{R}_{\geq0}$ the maximum capacity of line $j$. For each $i\in\mathcal{V}_l$, denote the $i$-th column of $H_l$ by $H_l(i)$.

For each $i\in\mathcal{V}_l$, we choose $\Lambda_i(L_i-R_i)=-\frac{1}{2}a_i(L_i-R_i)^2+b_i(L_i-R_i)$ with $a_i>0$. For the pricing policy $\Upsilon$, we adopt (4) in \cite{Bulow.Peiderer:83}, i.e., $\Upsilon(y)=\lambda y^{\frac{1}{\tau}}$ with $\lambda>0$ and $\tau\in(0,1)$. To make $\frac{1}{\tau}$ an integer, we choose $\tau=0.5$.

\emph{$\bullet$ Derivation of update rule \eqref{SCP of DC}.} By first introducing Lagrange multipliers to deal with the constraints of \eqref{demand response} and then applying the projected gradient method (in both primal and dual spaces), we can derive the update rule in the form of problem \eqref{SCP of DC} as follows:
\begin{align*}
&R_i(k+1)=\mathbb{P}_{[0,L_i]}[R_i(k)-\gamma(k)(c_i-a_i(L_i-R_i(k))\nonumber\\
&\qquad\qquad\quad\,+b_i-2\lambda(\textbf{1}_{|\mathcal{V}_l|}^T(L-R(k)))(L_i-R_i(k))\nonumber\\
&\qquad\qquad\quad\,-\lambda(\textbf{1}_{|\mathcal{V}_l|}^T(L-R(k)))^2-\mu_0(k)\nonumber\\
&\qquad\qquad\quad\,+\mu_+^T(k)H_l(i)-\mu_-^T(k)H_l(i))],\;\;\forall i\in\mathcal{V}_l\\
&\mu_0(k+1)=\mathbb{P}_{\mathbb{R}_{\geq0}}[\mu_0(k)+\gamma(k)(\textbf{1}_{|\mathcal{V}_l|}^T(L-R(k))\nonumber\\
&\qquad\qquad\quad\,-\textbf{1}_{|\mathcal{V}_s|}^TS)]\\
&\mu_+(k+1)=\mathbb{P}_{\mathbb{R}_{\geq0}^{|\mathcal{E}_p|}}[\mu_+(k)+\gamma(k)(H_sS-H_l(L-R)\nonumber\\
&\qquad\qquad\quad\,-f^{\rm max})]\\
&\mu_-(k+1)=\mathbb{P}_{\mathbb{R}_{\geq0}^{|\mathcal{E}_p|}}[\mu_-(k)+\gamma(k)(-H_sS+H_l(L-R)\nonumber\\
&\qquad\qquad\quad\,-f^{\rm max})]
\end{align*}
where $\mu_0\in\mathbb{R}_{\geq0}$, $\mu_+\in\mathbb{R}_{\geq0}^{|\mathcal{E}_p|}$ and $\mu_-\in\mathbb{R}_{\geq0}^{|\mathcal{E}_p|}$ are dual variables associated with the first, second and third constraint of \eqref{demand response}, respectively. Notice that $(\mu_0,\mu_+,\mu_-)$ are global dual variables since the constraints of \eqref{demand response} are shared by all the customers. We equally partition $(\mu_0,\mu_+,\mu_-)$ into $|\mathcal{V}_l|$ parts and each customer $i$ holds and updates one part of the overall dual states, denoted by $\mu_i$. Such a partition on the one hand reduces the computational burden (caused by encryption and decryption of the dual variables) of each customer, and on the other hand reduces the amount of information that can be accessed by each customer (since each customer only holds one subset of the dual variable update rule functions).

\emph{$\bullet$ Privacy issue.} In the demand response problem \eqref{demand response}, the parameters $(\lambda,H_s,H_l,S,f^{\rm max})$ are held by and private to the utility company (the system operator). On the other hand, each customer $i$ holds its primal and dual variables $(R_i,\mu_i)$ and coefficients $(a_i,b_i,c_i,L_i)$, whose values are private to customer $i$.

\subsubsection{Simulation results}

\begin{figure}
\begin{center}
\includegraphics[width=0.8\linewidth]{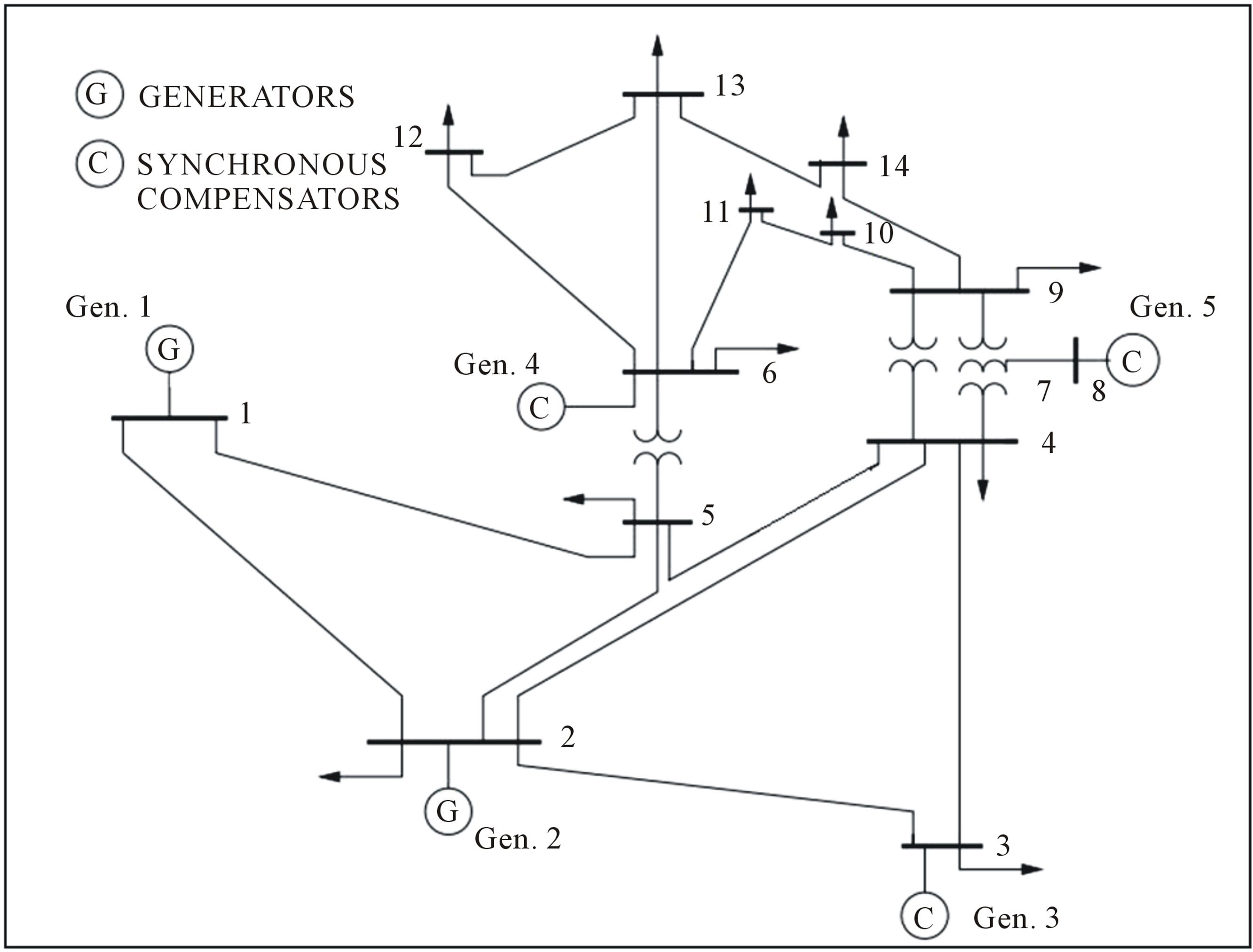}
\caption{IEEE 14-bus Test System}
\label{IEEE14}
\end{center}
\end{figure}

We use the IEEE 14-bus Test System \cite{IEEE14buspaper:1973}, shown by Fig. \ref{IEEE14}, as the demand response network. There are two generators and eleven loads in the system. The system parameters are adopted from MATPOWER \cite{RDZ-CEMS-RJT:2011}. In the simulation, we tested different lengths of keys and have checked that Assumption \ref{asm: key large enough} is satisfied for all the cases.

\begin{figure}
\begin{center}
\includegraphics[width=1\linewidth]{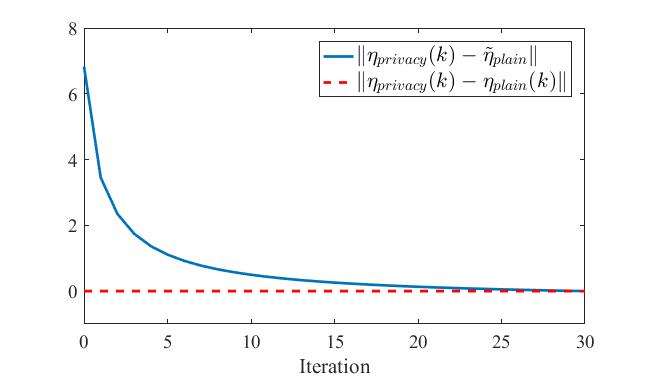}
\caption{Trajectories of $\|\eta_{privcy}(k)-\tilde \eta_{plain}\|$ and $\|\eta_{privcy}(k)-\eta_{plain}(k)\|$}
\label{DRP_err}
\end{center}
\end{figure}

\emph{$\bullet$ Perfect correctness.} We stack the collective primal and dual variables $(R,\mu_0,\mu_+,$ $\mu_-)$ into a single column vector, denoted by $\eta$. To show the correctness of Algorithm \ref{algo: ITS DOP}, we first obtain a benchmark evolution sequence of $\eta$ by simulating the plain projected gradient method without applying the homomorphic encryption mechanism of Algorithm \ref{algo: ITS DOP}. The precision level is set as keeping four decimal fraction digits during the whole process. The derived benchmark sequence of $\eta$ is denoted by $\{\eta_{plain}(k)\}$ and the state of the final iteration (iteration 30) is treated as the benchmark equilibrium, denoted by $\tilde\eta_{plain}$. We next simulate Algorithm \ref{algo: ITS DOP} with the same initial states and precision level as those of the benchmark simulation and the evolution sequence of $\eta$ is denoted by $\{\eta_{privcy}(k)\}$. The trajectories of $\|\eta_{privcy}(k)-\tilde \eta_{plain}\|$ and $\|\eta_{privcy}(k)-\eta_{plain}(k)\|$ are shown in Fig. \ref{DRP_err}. In Fig. \ref{DRP_err}, the trajectory of $\|\eta_{privcy}(k)-\tilde \eta_{plain}\|$ (blue solid line) shows the convergence behavior of $\{\eta_{privcy}(k)\}$ and the trajectory of $\|\eta_{privcy}(k)-\eta_{plain}(k)\|$ (red dashed line), which is constant at $0$, shows that $\eta_{privcy}(k)$ is exactly equal to $\eta_{plain}(k)$ at all iterations which verifies the prefect correctness of Algorithm \ref{algo: ITS DOP}.

\emph{$\bullet$ Computational efficiency.} We simulate with different key lengths to study the efficiency of the algorithm. The running time is shown by Table \ref{DRP time table}. The first column is the length of the private key $w$ in bit and the second (resp. third) column is the average (resp. maximum) time per iteration per customer in second, which consists of a single customer's encryption, encrypted computation, decryption, transformation between real numbers and integers and local update for a single iteration. Table \ref{DRP time table} shows that the proposed private key secure computation algorithm can be efficiently implemented.

\begin{table}[!t]
\renewcommand{\arraystretch}{1}
\caption{Running time of Algorithm \ref{algo: ITS DOP} on the demand response problem}
\label{DRP time table}
\centering
\begin{tabular}{c|c|c}
\hline
Key length & average time & maximum time\\
(bit) & /iter./customer (s) & /iter./customer (s)\\
\hline
500 & 0.0097 & 0.0099\\
\hline
1000 & 0.0100 & 0.0102\\
\hline
2000 & 0.0101 & 0.0103\\
\hline
3000 & 0.0104 & 0.0105\\
\hline
4000 & 0.0110 & 0.0112\\
\hline
\end{tabular}
\end{table}

\subsection{Algorithm \ref{algo: semantically secure DOP}}

We next simulate Algorithm \ref{algo: semantically secure DOP} by an optimal power flow (OPF) problem.

\subsubsection{Optimal power flow problem}

\emph{$\bullet$ Problem formulation.} Consider a power network comprising a set of agents $\mathcal{V} \buildrel \Delta \over = \{1,\cdots,N\}$ and a system operator. In an OPF setting, the agents could be power generators who supply electric energy via reference buses and the system operator could be an energy manager responsible for power supply regulation. For each generator $i$, let $\mathcal{N}_i\subseteq \mathcal{V}$ identify the set of neighbors to generator $i$. We adopt the OPF model from page 514 of \cite{Wood:1996} with the simplification that each bus has only one generator and one load. The physical meanings of the parameters and variables of the OPF problem are listed in Table \ref{OPF table}. The OPF problem is formulated as follows:
\begin{align}
\label{OPF}
&\!\!\!\!\!\!\!\!\!\!\!\mathop {\min }\nolimits_{P\in[\underline P,\overline P],\theta \in {\mathbb{R}^{N}}} \sum\nolimits_{i \in \mathcal{V}} {(a_iP_i^2+b_iP_i)}\nonumber\\
{\rm{s.t.}}\,\,&{P_{{i}}} - L_i = D_i\tilde\omega+\sum\nolimits_{j \in \mathcal{N}_i} t_{ij}(\theta_i-\theta_j) ,\,\forall i\in\mathcal{V},\nonumber\\
&t_{ij}(\theta_i-\theta_j) \le {\overline {P}_{ij}},\,\forall j \in \mathcal{N}_i,\,\forall i \in \mathcal{V}
\end{align}
where $a_i>0$ for all $i\in\mathcal{V}$, $P=(P_{i})\in\mathbb{R}^N$, $\theta=(\theta_{i})\in\mathbb{R}^N$, and $\tilde\omega=60{\rm Hz}$ is the steady state of $\omega$. In problem \eqref{OPF}, the objective function measures the power losses or the cost of supplied power. The equality constraints represent the power balance between neighboring reference buses. The inequality constraints depict the capacity limit of a line connecting two neighboring buses.

\begin{table}[!t]
\renewcommand{\arraystretch}{1}
\caption{Parameters/variables of the OPF problem}
\label{OPF table}
\centering
\begin{tabular}{|c|c|}
\hline
$a,b$ & cost parameters\\
\hline
$\omega$ & angular frequency\\
\hline
$\theta$ & phase angle\\
\hline
$P$ & mechanical power\\
\hline
$\underline P$ & lower limit of mechanical power\\
\hline
$\overline P$ & upper limit of mechanical power\\
\hline
$L$ & power load\\
\hline
$\overline P_{ij}$ & capacity of line connecting buses $i$ and $j$\\
\hline
$D$ & load-damping constant\\
\hline
$t_{ij}$ & tie-line stiffness coefficient\\
\hline
\end{tabular}
\end{table}

Notice that each summand of the objective function of \eqref{OPF}, $a_iP_i^2+b_iP_i$, only depends on and is convex in generator $i$'s local state $P_{i}$ and the constraints of problem \eqref{OPF} are affine in $(P_i,\theta_i)$. Thus, problem \eqref{OPF} is separable and equivalent to the distributed optimization problem where, given $(P_{{-i}},\theta_{-i})$, each generator $i$ aims to solve the following optimization problem:
\begin{align}
\label{separated OPF}
&\!\!\!\!\!\!\!\!\!\!\!\mathop {\min }\nolimits_{P_i\in[\underline P_i,\overline P_i],\theta_i \in {\mathbb{R}}}\,\, a_iP_i^2+b_iP_i\nonumber\\
{\rm{s.t.}}\,\,&{P_{{\ell}}} - L_\ell = D_\ell\tilde\omega+\sum\limits_{j \in \mathcal{N}_\ell} t_{\ell j}(\theta_\ell-\theta_j) ,\;\forall \ell\in\mathcal{N}_i\cup\{i\},\nonumber\\
&\!\!\!\!\!t_{ij}(\theta_i-\theta_j) \le {\overline {P}_{ij}},\,\,t_{ji}(\theta_j-\theta_i) \le {\overline {P}_{ji}},\;\forall j \in \mathcal{N}_i.
\end{align}

\emph{$\bullet$ Derivation of update rule \eqref{SCP of DC}.} Denote by $\lambda_\ell$ for $\ell\in\mathcal{N}_i\cup\{i\}$, $\mu_{ij}$ and $\mu_{ji}$ for $j\in\mathcal{N}_i$ the dual variables associated with the set of equality constraints, the first set of inequality constraints and the second set of inequality constraints of \eqref{separated OPF}, respectively. To reduce the computational burden and the accessible amount of information of each generator, we let each generator $i$ hold and update $(P_i,\theta_i,\lambda_i,\mu_{ij},j\in\mathcal{N}_i)$ (so that $\lambda_j$ and $\mu_{ji}$ for each $j\in\mathcal{N}_i$ are not held or updated by generator $i$, but by generator $j$). By introducing Lagrange multipliers and applying the projected gradient method, we can derive the update rule in the form of problem \eqref{SCP of DC} as follows:
\begin{align*}
&P_i(k+1)=\mathbb{P}_{[\underline P_i,\overline P_i]}[P_i(k)-\gamma(k)(2a_iP_i(k)+b_i\nonumber\\
&\qquad\qquad\quad\,-\lambda_i(k))]\\
&\theta_i(k+1)=\theta_i(k)-\gamma(k)\sum\nolimits_{j\in\mathcal{N}_i}(t_{ij}(\lambda_i(k)+\mu_{ij}(k))\nonumber\\
&\qquad\qquad\quad\,-t_{ij}(\lambda_j(k)+\mu_{ji}(k)))\\
&\lambda_i(k+1)=\lambda_i(k)+\gamma(k)(L_i-P_i(k)+D_i\tilde\omega\nonumber\\
&\qquad\qquad\quad\,+\sum\nolimits_{j\in\mathcal{N}_i}t_{ij}(\theta_i(k)-\theta_j(k)))\\
&\mu_{ij}(k+1)=\mathbb{P}_{\mathbb{R}_{\geq0}}[\mu_{ij}(k)+\gamma(k)(t_{ij}(\theta_i-\theta_j)-{\overline {P}_{ij}})].
\end{align*}

\emph{$\bullet$ Privacy issue.} Each generator $i$ holds its primal and dual variables $(P_i,\theta_i,\lambda_i,\mu_{ij},j\in\mathcal{N}_i)$ and the coefficients $(a_i,b_i,L_i,\overline P_i,\underline P_i)$, whose values should be kept private to generator $i$. Otherwise, an adversary who knows the current power system state could infer a critical contingency and launch a targeted attack to implement it \cite{GD-RBB-GG-RHC:2013}. On the other hand, the energy manager (the system operator) holds the line-dependent coefficients $(\overline P_{ij},D_i,t_{ij})$ whose values should be kept private to the energy manager. In power systems, the line-dependent coefficients are often very important to the system operator for security consideration. Leaks of these coefficients could be financially damaging and cause potential threat to national security \cite{ARB-DKM-BCL-PR:2013}, \cite{ARB-DKM-PR-BCL:2012}. For example, an adversary could leverage such information to infer expected congestion in the power system and use this knowledge for insider trading in the power markets \cite{GD-RBB-GG-RHC:2013}.

\subsubsection{Simulation results}

\begin{figure}
\begin{center}
\includegraphics[width=0.5\linewidth]{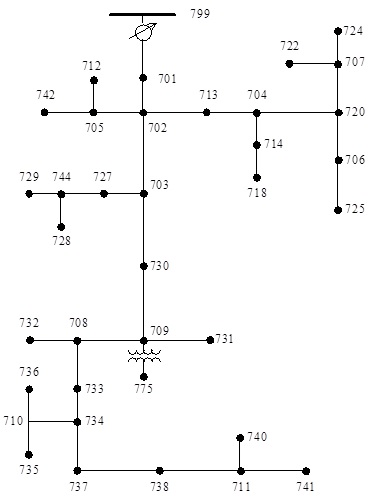}
\caption{IEEE 37-bus Test System}
\label{IEEE37}
\end{center}
\end{figure}

We use the IEEE 37-bus Test System \cite{IEEE-1991}, shown by Fig. \ref{IEEE37}, as the OPF network. The values of the parameters are chosen as follows: for all $i\in\mathcal{V}$, $D_i=1MW/Hz$, $t_{ij}=1.5MW/rad$, $a_i=0.1$, $b_i=10$, $\overline P_{i}=100MW$, $\underline P_{i}=10MW$, $P_{L_i}=10MW$ and $\overline P_{ij}=80MW$. We tested different lengths of keys and have checked that Assumption \ref{asm: public key large enough} is satisfied for all the tested cases.

\begin{figure}
\begin{center}
\includegraphics[width=1\linewidth]{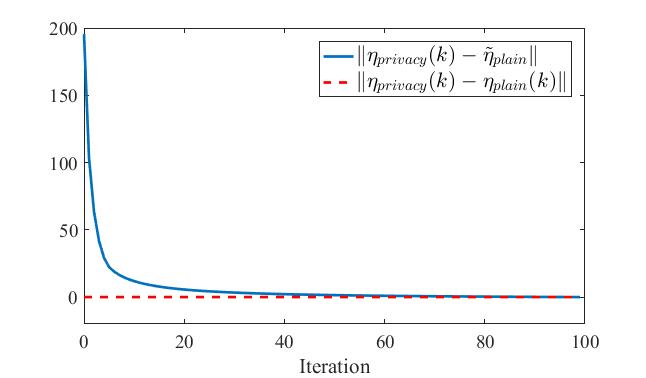}
\caption{Trajectories of $\|\eta_{privacy}(k)-\tilde \eta_{plain}\|$ and $\|\eta_{privacy}(k)-\eta_{plain}(k)\|$}
\label{OPF_err}
\end{center}
\end{figure}

\emph{$\bullet$ Perfect correctness.} We stack the collective primal and dual variables $(P,\theta,\lambda,\mu)$ into a single column vector $\eta$. Similar simulation process of Algorithm \ref{algo: ITS DOP} is performed here. Fig. \ref{OPF_err} shows the trajectories of $\|\eta_{privacy}(k)-\tilde \eta_{plain}\|$ and $\|\eta_{privacy}(k)-\eta_{plain}(k)\|$, which verify the perfect correctness of Algorithm \ref{algo: semantically secure DOP}.

\emph{$\bullet$ Computational efficiency.} The running time is shown by Table \ref{OPF time table}. The first column is the length of the public key $\alpha_i$'s in bit. The average and maximum time per iteration per generator in the second and third columns consists of a single generator's encryption, encrypted computation, decryption, transformation between real numbers and integers and local update for a single iteration. Table \ref{DRP time table} shows that the proposed public key secure computation algorithm can be efficiently implemented.

\begin{table}[!t]
\renewcommand{\arraystretch}{1}
\caption{Running time of Algorithm \ref{algo: semantically secure DOP} on the OPF problem}
\label{OPF time table}
\centering
\begin{tabular}{c|c|c}
\hline
Key length & average time & maximum time\\
(bit) & /iter./generator (s) & /iter./generator (s)\\
\hline
500 & 0.0147 & 0.0193\\
\hline
1000 & 0.0424 & 0.0461\\
\hline
2000 & 0.2168 & 0.2492\\
\hline
3000 & 0.6825 & 0.7461\\
\hline
4000 & 1.5487 & 1.7259\\
\hline
\end{tabular}
\end{table}

\section{Proofs\label{section proofs}}

In this section, we provide the proofs of the theoretical results derived in the previous sections.

\subsection{Proof of Theorem \ref{theorem: GNEP}\label{proof of GNEP theorem}}

\begin{pf}
1) Proof of correctness.

We first fix a step $k$ and some $I\in\mathcal{V}$ and $J\in\{1,\cdots,n_I\}$ to show that $\hat\Phi_{IJ}^s(k)=\Phi_{IJ}(x(k))$. Recall that $10^\sigma x(k)$ and $10^\sigma c$ are both integer vectors.

For each $\ell=1,\cdots,\kappa_{IJ}$, $Q_{IJ}^\ell(\hat x(k),\hat y)$ is a monomial evaluated at $(\hat x(k),\hat y)$. Since $\hat y_{\rm SO}=10^\sigma c_{\rm SO}$ and, for each $i\in\mathcal{V}$, $\hat y_i=u_{y_i}w+10^{\sigma}c_i$ and $\hat x_i(k)=u_{x_i}(k)w+10^{\sigma}x_i(k)$, $Q_{IJ}^\ell(\hat x(k),\hat y)$ can be expressed as
\begin{align}
\label{polynamial proof step 1}
\!\!\!\!\!Q_{IJ}^\ell(\hat x(k),\hat y)=\psi_{IJ}^\ell(k)w+10^{{\rm deg}(Q_{IJ}^\ell)\sigma}Q_{IJ}^\ell(x(k),c)
\end{align}
where $\psi_{IJ}^\ell(k)$ is an integer. To see this, consider the example $Q_{IJ}^\ell(x,y)=y_Ix_I$, where $y_I$ and $x_I$ are assumed to be scalars for simplicity. Then, ${\rm deg}(Q_{IJ}^\ell)=2$. We have
\begin{align*}
&Q_{IJ}^\ell(\hat x(k),\hat y)=\hat y_I\hat x_I(k)\\
&=(u_{y_I}w+10^{\sigma} c_I)(u_{x_I}(k)w+10^{\sigma} x_I(k))\\
&=(u_{y_I}u_{x_I}(k)w+10^{\sigma}(x_I(k)u_{y_I}+c_Iu_{x_I}(k)))w\\
&\quad+10^{2\sigma}c_Ix_I(k)=\psi_{IJ}^\ell(k) w+10^{{\rm deg}(Q_{IJ}^\ell)\sigma}Q_{IJ}^\ell(x(k),c)
\end{align*}
where $\psi_{IJ}^\ell(k)=u_{y_I}u_{x_I}(k)w+10^{\sigma}(x_I(k)u_{y_I}+c_Iu_{x_I}(k))$. By \eqref{polynamial proof step 1}, for each $\ell=1,\cdots,\kappa_{IJ}$, we then have
\begin{align}
\label{polynamial proof step 2}
&10^{({\rm deg}(\Phi_{IJ}^s)-{\rm deg}(Q_{IJ}^\ell))\sigma}Q_{IJ}^\ell(\hat x(k),\hat y)\nonumber\\
&=\rho_{IJ}^\ell(k) w+10^{{\rm deg}(\Phi_{IJ}^s)\sigma}Q_{IJ}^\ell(x(k),c)
\end{align}
where $\rho_{IJ}^\ell(k)=10^{({\rm deg}(\Phi_{IJ}^s)-{\rm deg}(Q_{IJ}^\ell))\sigma}\psi_{IJ}^\ell(k)$ is an integer. By \eqref{polynamial proof step 2}, we can further obtain
\begin{align}
\label{polynamial proof step 3}
&\bar \Phi_{IJ}^s(k)=\sum\nolimits_{\ell=1}^{\kappa_{IJ}}[10^{({\rm deg}(\Phi_{IJ}^s)-{\rm deg}(Q_{IJ}^\ell))\sigma}Q_{IJ}^\ell(\hat x(k),\hat y)]\nonumber\\
&=w\sum\nolimits_{\ell=1}^{\kappa_{IJ}}{\rho_{IJ}^\ell(k)}+10^{{\rm deg}(\Phi_{IJ}^s)\sigma}\sum\nolimits_{\ell=1}^{\kappa_{IJ}}Q_{IJ}^\ell(x(k),c)\nonumber\\
&=\varsigma_{IJ}(k) w+10^{{\rm deg}(\Phi_{IJ}^s)\sigma} \Phi_{IJ}(x(k))
\end{align}
where $\varsigma_{IJ}(k)=\sum\nolimits_{\ell=1}^{\kappa_{IJ}}{\rho_{IJ}^\ell(k)}$ is an integer. By \eqref{polynamial proof step 3}
\begin{align}
\label{polynamial proof step 4}
\bar \Phi_{IJ}^s(k)\!\!\mod w=10^{{\rm deg}(\Phi_{IJ}^s)\sigma} \Phi_{IJ}(x(k))\!\!\mod w.
\end{align}
By Assumption \ref{asm: key large enough}, $|10^{{\rm deg}(\Phi_{IJ}^s)\sigma}\Phi_{IJ}(x(k))|<{(w-1)}/{2}$. By Lemma \ref{T function property}, we then have $\hat\Phi_{IJ}^s(k)=T_{{\rm deg}(\Phi_{IJ}^s)\sigma,w}$ $(10^{{\rm deg}(\Phi_{IJ}^s)\sigma} \Phi_{IJ}(x(k))\mod w)=\Phi_{IJ}(x(k))$. The above analysis holds for $\Phi_{ij}(x(k))$ for any $k$ and $(i,j)$. Hence, for each $k$ and $i\in\mathcal{V}$, we have $\hat\Phi_i^s(k)=\Phi_i(x(k))$.

2) Proof of security.

The security proof adopts the simulation paradigm which is a standard technique for proving SMC \cite{RC-ID-JBN:2015}, \cite{YL-BP:09}. First we show by Definition \ref{secure compute def} that Algorithm \ref{algo: ITS DOP} securely computes $\{\Phi_1(x(k)),\cdots,\Phi_N(x(k))\}$ between the agents. During each step $k$ of the execution of Algorithm \ref{algo: ITS DOP}, for each $i\in\mathcal{V}$, agent $i$ only receives $\bar\Phi_i^s(k)$. Thus, the view of agent $i$ during step $k$ of Algorithm \ref{algo: ITS DOP} is ${\rm VIEW}_i(k)=\{x_i(k),C_\Phi^i,X_i,\sigma,w,u_{x_i}(k),u_{y_i},\bar\Phi_i^s(k),$ $\Phi_i(x(k))\}$, where $x_i(k)$, $C_\Phi^i$, $X_i$, $\sigma$, $w$, $u_{x_i}(k)$ and $u_{y_i}$ are inputs of agent $i$ and $\Phi_i(x(k))$ is agent $i$'s output which can be inferred from $\bar\Phi_i^s(k)$ by the correctness analysis. We need to construct an algorithm $S$ by which, for each $i\in\mathcal{V}$, given agent $i$'s inputs $\{x_i(k),C_\Phi^i,X_i,\sigma,w,u_{x_i}(k),u_{y_i}\}$ and output $\Phi_i(x(k))$, agent $i$ can generate a set of data $S(i,k)$ such that $S(i,k)\mathop  \equiv \limits^{c}{\rm VIEW}_i(k)$. Agent $i$ then only has to simulate $\bar\Phi_i^s(k)$, i.e., to generate $\bar\Phi_{i}^s(k)'$ by $S$ such that $\{\bar\Phi_{i}^s(k)'\}\mathop  \equiv \limits^{c}\{\bar\Phi_{i}^s(k)\}$. By \eqref{polynamial proof step 3}, we have $\bar\Phi_{ij}^s(k)=\varsigma_{ij}(k) w+10^{{\rm deg}(\Phi_{ij}^s)\sigma} \Phi_{ij}(x(k))$. Since each agent $\ell$ randomly chooses $u_{x_\ell}(k)$ and $u_{y_\ell}$, $\varsigma_{ij}(k)$ is a random number to agent $i$. Let $\bar c_i$ and $\bar c_{-i}$ be the vectors of the elements of $C_\Phi^i$ and $C_\Phi\backslash C_\Phi^i$, respectively. Agent $i$ simulates $\bar\Phi_i^s(k)$ as follows. First, agent $i$ generates $x_{-i}(k)'$ and $\bar c_{-i}'$ such that $\Phi_i^s(x_i(k),x_{-i}(k)',\bar c_i,\bar c_{-i}')=\Phi_i(x(k))$. Then, agent $i$ randomly chooses $(u_{x_{-i}}(k)',u_{y_{-i}}')$ following the same distribution as $(u_{x_{-i}}(k),u_{y_{-i}})$ and computes $\hat x_i(k)=u_{x_i}(k)w+x_i(k)$, $\hat y_i=u_{y_i}w+\bar c_i$, $\hat x_{-i}(k)'=u_{x_{-i}}(k)'w+x_{-i}(k)'$ and $\hat y_{-i}'=u_{y_{-i}}'w+\bar c_{-i}'$. Agent $i$ then computes $\bar\Phi_{i}^s(k)'$ in the same way as that of line 6 of Algorithm \ref{algo: ITS DOP} by using $\{\hat x_i(k),\hat x_{-i}(k)',\hat y_i,\hat y_{-i}'\}$. By the same arguments in the correctness analysis part, we have $\bar\Phi_{ij}^s(k)'=\varsigma_{ij}(k)' w+10^{{\rm deg}(\Phi_{ij}^s)\sigma} \Phi_{ij}(x(k))$, where $\varsigma_{ij}(k)'$ is an integer which has the same distribution with $\varsigma_{ij}(k)$. Thus, we have $\{\bar\Phi_{i}^s(k)'\}\mathop  \equiv \limits^{c}\{\bar\Phi_{i}^s(k)\}$. By Definition \ref{secure compute def}, Algorithm \ref{algo: ITS DOP} securely computes $\{\Phi_1(x(k)),\cdots,\Phi_N(x(k))\}$ between the agents.

Next we show that Algorithm \ref{algo: ITS DOP} is as hard as the approximate GCD problem for the system operator to solve $x(k)$ and $c_{\mathcal{V}}$ at each step $k$. Under Assumption \ref{asm: one shot}, at each step $k$, the system operator knows $\hat x(k)=u_x(k)w+10^\sigma x(k)$ and $\hat y=u_yw+10^\sigma c_{\mathcal{V}}$. To infer $x(k)$ and $c_{\mathcal{V}}$ from $\hat x(k)$ and $\hat y$, the system operator has to recover the value of $w$, which is exactly the approximate GCD problem.
\end{pf}

\subsection{Proof of Theorem \ref{theorem: DOP special case}\label{proof of DOP special case theorem}}

\begin{pf}
1) Proof of correctness.

For each step $k$ and compatible $(i,\ell)$, since $\bar \Phi_{i\ell}^s(k)=(\beta_i^{10^{2\sigma}B_{i\ell}\mod \alpha_i}\prod\nolimits_{j=1}^N\prod\nolimits_{v=1}^{n_j}\hat x_{jv}^i(k)^{10^{\sigma}A_{i\ell}^{jv}\mod \alpha_i}) \mod$ $\alpha_i^2$, by (3) of Theorem \ref{theorem: Paillier}, we have $\frac{(\bar\Phi_{i\ell}^s(k)^{\nu} \mod \alpha_i^2)-1}{\alpha_i}\cdot \pi_i \mod \alpha_i=10^{2\sigma}(A_{i\ell}x(k)+B_{i\ell})\mod \alpha_i$. By Assumption \ref{asm: public key large enough}, $10^{2\sigma}|A_{i\ell}x(k)+B_{i\ell}|\leq(\alpha_i-1)/2$. By Lemma \ref{T function property}, we have $\hat\Phi_{i\ell}^s(k)=T_{2\sigma,\alpha_i}(10^{2\sigma}(A_{i\ell}x(k)+B_{i\ell})\mod \alpha_i)=A_{i\ell}x(k)+B_{i\ell}=\Phi_{i\ell}(x(k))$. The above analysis holds for any step $k$ and any pair $(i,\ell)$. Hence, for each step $k$ and each $i\in\mathcal{V}$, we have $\hat\Phi_i^s(k)=\Phi_i(x(k))$.

2) Proof of security.

We first show by Definition \ref{secure compute def} that Algorithm \ref{algo: semantically secure DOP} securely computes $\{\Phi_1(x(k)),\cdots,\Phi_N(x(k))\}$ between the agents. During each step $k$ of the execution of Algorithm \ref{algo: semantically secure DOP}, for each $i\in\mathcal{V}$, agent $i$ only receives $\bar\Phi_i^s(k)$ and ${\rm VIEW}_i(k)=\{x_i(k),\sigma,\alpha_i,\beta_i,\nu_i,\pi_i,\bar\Phi_i^s(k),\Phi_i(x(k))\}$. We need to construct an algorithm $S$ by which, given agent $i$'s inputs $\{x_i(k),\sigma,\alpha_i,\beta_i,\nu_i,\pi_i\}$ and output $\Phi_i(x(k))$, agent $i$ can generate a set of data $S(i,k)$ such that $S(i,k)\mathop  \equiv \limits^{c}{\rm VIEW}_i(k)$. Again, agent $i$ only has to generate $\bar\Phi_{i}^s(k)'$ by $S$ such that $\{\bar\Phi_{i}^s(k)'\}\mathop  \equiv \limits^{c}\{\bar\Phi_{i}^s(k)\}$. Agent $i$ simulates $\bar\Phi_i^s(k)$ as follows. First, agent $i$ generates $x_{-i}(k)'$ and $(A_i',B_i')$ such that $\Phi_{i}^s(x_i(k),x_{-i}(k)',A_i',B_i')$ $=\Phi_i(x(k))$. Then, for each $j\in\mathcal{V}\backslash\{i\}$, agent $i$ randomly chooses $r_j(k)'$ where each $r_{j\ell}(k)'\in\mathbb{Z}_{\alpha_i}^*$ and encrypts $x_j(k)'$ by line 4 of Algorithm \ref{algo: semantically secure DOP} to obtain $\hat x_j(k)'$. Then, agent $i$ computes $\bar\Phi_i^s(k)'$ in the same way as line 5 of Algorithm \ref{algo: semantically secure DOP} by using $(\hat x_i(k),\hat x_{-i}(k)',A_i',B_i')$. For each $j=1,\cdots,n_i$, by (3) of Theorem \ref{theorem: Paillier}, we have that $\bar\Phi_{ij}^s(k)'$ satisfies the decryption correctness, i.e., $\frac{(\bar\Phi_{ij}^s(k)'^{\nu} \mod \alpha_i^2)-1}{\alpha_i}\cdot \pi_i \mod \alpha_i=10^{2\sigma}\Phi_{ij}(x(k))\mod \alpha_i$. Hence, $\{\bar\Phi_{i}^s(k)'\}\mathop  \equiv \limits^{c}\{\bar\Phi_{i}^s(k)\}$. By Definition \ref{secure compute def}, Algorithm \ref{algo: semantically secure DOP} securely computes $\{\Phi_1(x(k)),\cdots,\Phi_N(x(k))\}$ between the agents.

We next show that Algorithm \ref{algo: semantically secure DOP} is semantically secure against the system operator. During the execution of Algorithm \ref{algo: semantically secure DOP}, at each step $k$, the system operator receives $N$ encrypted vectors of $x(k)$, each one encrypted by a different set of Paillier keys. By Theorem \ref{theorem: Paillier}, the Paillier encryption scheme is semantically secure for encrypting a plaintext a single time. Then, by Theorem 5.2.10 of \cite{OG:2004}, the Paillier encryption scheme is also semantically secure for encrypting a message $N$ times by different keys against an adversary who knows all the $N$ ciphertexts. Denote by $E(\cdot)$ the encryption operator of Algorithm \ref{algo: semantically secure DOP}. By Definition \ref{def encryption semantic security}, to the system operator, $E(x)\mathop  \equiv \limits^cE(x')$ for any plaintexts $x,x'$ even if the system operator knows $x$ and $x'$. Thus, to the system operator, $E(x(k))\mathop \equiv \limits^cE(x(k+\kappa))$ for any $k$ and any $\kappa$ even if the system operator knows $x(k)$ and $x(k+\kappa)$ or any relation between $x(k)$ and $x(k+\kappa)$. Hence, even if the system operator knows the update rule of $x(k)$ and launchs causal attacks, Algorithm \ref{algo: semantically secure DOP} is still semantically secure. This establishes the semantic security of Algorithm \ref{algo: semantically secure DOP}.
\end{pf}

\subsection{Proof of Theorem \ref{IO resist linear case I}\label{proof of IO resist linear case I}}

\begin{pf}
We focus on an adversary $i\in\mathcal{V}$. By Definition \ref{secure from output}, we are to show that agent $i$ cannot uniquely determine $(\{x_{-i}(0),\cdots,x_{-i}(K)\},X_{-i},[B_{j\ell}]_{(j,\ell)\in\Omega_i'})$ and the uncertainty is unbounded for any $K\in\mathbb{N}$. For each $j\in\mathcal{V}$ and $\ell=1,\cdots,n_j$, denote the true $(X_{j\ell},B_{j\ell},\{x_{j\ell}(0),\cdots,x_{j\ell}(K)\})$ by $(\bar X_{j\ell},\bar B_{j\ell},\{\bar x_{j\ell}(0),$ $\cdots,\bar x_{j\ell}(K)\})$. Up to step $k$, agent $i$ knows $\bar X_i$, $H_{j\ell}$ and $A_{j\ell}$ for all $j\in\mathcal{V}$ and $\ell\in\{1,\cdots,n_j\}$, $B_{j\ell}$ for $(j,\ell)\in\Omega_i$, the sequences $\{\bar x_i(0),\cdots,\bar x_i(k)\}$, $\{\Phi_i(\bar x(0)),\cdots,\Phi_i(\bar x(k))\}$ and $\{\gamma(0),\cdots,\gamma(k)\}$. For convenience of notation, let $\hat x_i(k)=\bar x_i(k)$ for all $k\in\{0,\cdots,K\}$. We aim to find $(\hat X_{-i},\{\hat x_{-i}(0),\cdots,\hat x_{-i}(K)\},$ $[\hat B_{j\ell}]_{(j,\ell)\in\Omega_i'})$ which satisfy the following constraints:
\begin{align}
\label{IO resist case I eq1_1}
&\hat x_{j\ell}(k+1)=\mathbb{P}_{\hat X_{j\ell}}[\hat x_{j\ell}(k)-\gamma(k)\hat x(k)^TH_{j\ell}\hat x(k)\nonumber\\
&+A_{j\ell}\hat x(k)+\bar B_{j\ell})],\;\forall (j,\ell)\in\Omega_i,\;j\neq i,\nonumber\\
&\ell\in\{1,\cdots,n_j\},\;k\in\{0,\cdots,K-1\},\\
\label{IO resist case I eq1_2}
&\hat x_{j\ell}(k+1)=\mathbb{P}_{\hat X_{j\ell}}[\hat x_{j\ell}(k)-\gamma(k)\hat x(k)^TH_{j\ell}\hat x(k)\nonumber\\
&+A_{j\ell}\hat x(k)+\hat B_{j\ell})],\;\forall (j,\ell)\in\Omega_i',\;j\neq i,\nonumber\\
&\ell\in\{1,\cdots,n_j\},\;k\in\{0,\cdots,K-1\},\\
\label{IO resist case I eq1_3}
&\!\!\!\!\!\!\left\{ {\begin{array}{*{20}{l}}
\Phi_{i\ell}(\bar x(k))=\hat x(k)^TH_{i\ell}\hat x(k)+A_{i\ell}\hat x(k)+\bar B_{i\ell},\\
\forall k\in\{0,\cdots,K\},\;\ell\in\{1,\cdots,n_i\},\;{\rm if}\;(i,\ell)\in\Omega_i,\\
\Phi_{i\ell}(\bar x(k))=\hat x(k)^TH_{i\ell}\hat x(k)+A_{i\ell}\hat x(k)+\hat B_{i\ell},\\
\forall k\in\{0,\cdots,K\},\;\ell\in\{1,\cdots,n_i\},\;{\rm if}\;(i,\ell)\in\Omega_i'.
\end{array}} \right.
\end{align}

Consider the following construction: $\hat x_j(k)=\bar x_j(k)+\delta_j$ for each $j\in\mathcal{V}$ and each $k\in\{0,\cdots,K\}$, where $\delta_i=0$, and for each $j\neq i$, $\delta_j\in\mathbb{R}^{n_j}$ is an arbitrary constant vector such that each entry of each $\delta_j$ is nonzero and $[(H^{-i})^T,(A^{-i})^T]^T\delta_{-i}=0$, with $\delta_{-i}=[\delta_v]_{v\neq i}$; $\hat X_j = \{x_j+\delta_j:x_j\in \bar X_j\}$ for each $j\neq i$; $\hat B_{j\ell}=\bar B_{j\ell}-A_{j\ell}\delta$ for each $(j,\ell)\in\Omega_i'$ with $\delta=[\delta_v]_{v\in\mathcal{V}}$. Assumption \ref{asm: input output inference} guarantees the existence of the above set of $\delta_j$'s. By the definition of $H^{-i}$ and $A^{-i}$ given in the fourth paragraph of Section \ref{observability section}, we have $(H_{j\ell}^{uv}+(H_{j\ell}^{vu})^T)\delta_v=0$ for all possible quadruples $(j,\ell,u,v)$, and $A_{j\ell}\delta=0$ for all $(j,\ell)\in\Omega_i$.

First we consider the case where $j\neq i$ and $(j,\ell)\in\Omega_i$. By this construction, for each step $k$, it holds that:
\begin{align}
\label{IO resist case III eq3}
&\mathbb{P}_{\hat X_{j\ell}}[\hat x_{j\ell}(k)-\gamma(k)(\hat x(k)^TH_{j\ell}\hat x(k)+A_{j\ell}\hat x(k)+\bar B_{j\ell})]\nonumber\\
&=\mathbb{P}_{\hat X_{j\ell}}[\bar x_{j\ell}(k)+\delta_{j\ell}-\gamma(k)(\sum\nolimits_{u,v=1}^N(\bar x_u(k)+\delta_u)^T\nonumber\\
&H_{j\ell}^{uv}(\bar x_v(k)+\delta_v)+\sum\nolimits_{v=1}^NA_{j\ell}^v(\bar x_v(k)+\delta_v)+\bar B_{j\ell})]\nonumber\\
&=\mathbb{P}_{\hat X_{j\ell}}[\bar x_{j\ell}(k)+\delta_{j\ell}-\gamma(k)(\bar x(k)^TH_{j\ell}\bar x(k)+A_{j\ell}\bar x(k)+\nonumber\\
&\bar B_{j\ell}+\sum\limits_{u,v=1}^N(\bar x_u(k)+\frac{1}{2}\delta_u)^T(H_{j\ell}^{uv}+(H_{j\ell}^{vu})^T)\delta_v+A_{j\ell}\delta)]\nonumber\\
&=\mathbb{P}_{\hat X_{j\ell}}[\bar x_{j\ell}(k)+\delta_{j\ell}-\gamma(k)(\bar x(k)^TH_{j\ell}\bar x(k)+A_{j\ell}\bar x(k)\nonumber\\
&+\bar B_{j\ell})]=\bar x_{j\ell}(k+1)+\delta_{j\ell}=\hat x_{j\ell}(k+1),\nonumber\\
&\forall (j,\ell)\in\Omega_i,\;j\neq i,\;\ell\in\{1,\cdots,n_j\}.
\end{align}

Next we consider the case where $j\neq i$ and $(j,\ell)\in\Omega_i'$. By the above construction, for each step $k$, it holds that:
\begin{align}
\label{IO resist case III eq4}
&\mathbb{P}_{\hat X_{j\ell}}[\hat x_{j\ell}(k)-\gamma(k)(\hat x(k)^TH_{j\ell}\hat x(k)+A_{j\ell}\hat x(k)+\hat B_{j\ell})]\nonumber\\
&=\mathbb{P}_{\hat X_{j\ell}}[\bar x_{j\ell}(k)+\delta_{j\ell}-\gamma(k)(\sum\nolimits_{u,v=1}^N(\bar x_u(k)+\delta_u)^TH_{j\ell}^{uv}\nonumber\\
&(\bar x_v(k)+\delta_v)+\sum\nolimits_{v=1}^NA_{j\ell}^v(\bar x_v(k)+\delta_v)+\bar B_{j\ell}-A_{j\ell}\delta)]\nonumber\\
&=\mathbb{P}_{\hat X_{j\ell}}[\bar x_{j\ell}(k)+\delta_{j\ell}-\gamma(k)(\bar x(k)^TH_{j\ell}\bar x(k)+A_{j\ell}\bar x(k)\nonumber\\
&+\bar B_{j\ell}+\sum\limits_{u,v=1}^N(\bar x_u(k)+\frac{1}{2}\delta_u)^T(H_{j\ell}^{uv}+(H_{j\ell}^{vu})^T)\delta_v)]\nonumber\\
&=\mathbb{P}_{\hat X_{j\ell}}[\bar x_{j\ell}(k)+\delta_{j\ell}-\gamma(k)(\bar x(k)^TH_{j\ell}\bar x(k)+A_{j\ell}\bar x(k)\nonumber\\
&+\bar B_{j\ell})]=\bar x_{j\ell}(k+1)+\delta_{j\ell}=\hat x_{j\ell}(k+1),\nonumber\\
&\forall (j,\ell)\in\Omega_i',\;j\neq i,\;\ell\in\{1,\cdots,n_j\}.
\end{align}

We next consider the case where $(i,\ell)\in\Omega_i$. By the above construction, for each step $k$, it holds that:
\begin{align}
\label{IO resist case III eq5}
&\hat x(k)^TH_{i\ell}\hat x(k)+A_{i\ell}\hat x(k)+\bar B_{i\ell}\nonumber\\
&=(\bar x(k)+\delta)^TH_{i\ell}(\bar x(k)+\delta)+A_{i\ell}(\bar x(k)+\delta)+\bar B_{i\ell}\nonumber\\
&=\bar x(k)^TH_{i\ell}\bar x(k)+A_{i\ell}\bar x(k)+\bar B_{i\ell}\nonumber\\
&+\sum\nolimits_{u,v=1}^N(\bar x_u(k)+\frac{1}{2}\delta_u)^T(H_{i\ell}^{uv}+(H_{i\ell}^{vu})^T)\delta_v+A_{i\ell}\delta\nonumber\\
&=\bar x(k)^TH_{i\ell}\bar x(k)+A_{i\ell}\bar x(k)+\bar B_{i\ell}=\Phi_{i\ell}(\bar x(k)),\nonumber\\
&\forall\ell\in\{1,\cdots,n_j\},\;{\rm if}\;(i,\ell)\in\Omega_i.
\end{align}

Finally we consider the case where $(i,\ell)\in\Omega_i'$. By the above construction, for each step $k$, it holds that:
\begin{align}
\label{IO resist case III eq6}
&\hat x(k)^TH_{i\ell}\hat x(k)+A_{i\ell}\hat x(k)+\hat B_{i\ell}\nonumber\\
&=(\bar x(k)+\delta)^TH_{i\ell}(\bar x(k)+\delta)+A_{i\ell}(\bar x(k)+\delta)+\bar B_{i\ell}\nonumber\\
&-A_{i\ell}\delta=\bar x(k)^TH_{i\ell}\bar x(k)+A_{i\ell}\bar x(k)+\bar B_{i\ell}\nonumber\\
&+\sum\nolimits_{u,v=1}^N(\bar x_u(k)+\frac{1}{2}\delta_u)^T(H_{i\ell}^{uv}+(H_{i\ell}^{vu})^T)\delta_v\nonumber\\
&=\bar x(k)^TH_{i\ell}\bar x(k)+A_{i\ell}\bar x(k)+\bar B_{i\ell}=\Phi_{i\ell}(\bar x(k)),\nonumber\\
&\forall\ell\in\{1,\cdots,n_j\},\;{\rm if}\;(i,\ell)\in\Omega_i'.
\end{align}

It is easy to see that \eqref{IO resist case III eq3} implies \eqref{IO resist case I eq1_1}, \eqref{IO resist case III eq4} implies \eqref{IO resist case I eq1_2}, and \eqref{IO resist case III eq5} and \eqref{IO resist case III eq6} imply \eqref{IO resist case I eq1_3}. Thus, the above construction of $(\{\hat x_{-i}(0),\cdots,\hat x_{-i}(K)\},\hat X_{-i},[\hat B_{j\ell}]_{(j,\ell)\in\Omega_i'})$ satisfies \eqref{IO resist case I eq1_1}, \eqref{IO resist case I eq1_2} and \eqref{IO resist case I eq1_3}. By the definition of $\Delta_j^i$ in Definition \ref{secure from output}, we have $(\{\hat x_j(0),\cdots,\hat x_j(K)\},\hat X_j,[\hat B_{j\ell}]_{(j,\ell)\in\Omega_i'})\in\Delta_j^i(\{\bar x(0),\cdots,\bar x(K)\},\bar X,\bar B)$ for each $j\neq i$. Fix any set of $\delta_j$'s constructed above. Then for an arbitrary nonzero real number $r$, $r\delta$ also satisfies the condition given in Assumption \ref{asm: input output inference}, i.e., $[(H^{-i})^T,(A^{-i})^T]^Tr\delta_{-i}=0$, and thus the construction of $(\{\hat x_j(0),\cdots,\hat x_j(K)\},\hat X_j,\hat B)$ by replacing $\delta_j$ with $r\delta_j$ also belongs to $\Delta_j^i(\{\bar x(0),\cdots,\bar x(K)\},$ $\bar X,\bar B)$. Notice that each entry of each $r\delta_j$ is nonzero and could be arbitrarily large by choosing $r$ arbitrarily large. Hence, by the definition of ${\rm Dist}_j^i$ in Definition \ref{secure from output}, for each $j\neq i$, we have ${\rm Dist}_j^i(\{\bar x(0),\cdots,\bar x(K)\},\bar X,\bar B)\geq\sup_{r\in\mathbb{R}}\|r\delta_j-\delta_j\|_2=\infty$. Since the above analysis holds for any $i\in\mathcal{V}$, we have $\min_{i\neq j}{\rm Dist}_j^i(\{\bar x(0),\cdots,\bar x(K)\},$ $\bar X,\bar B)=\infty$ for any $j\in\mathcal{V}$. By Definition \ref{secure from output}, the sequence of functions $\{\Phi(x(0)),\cdots,\Phi(x(K))\}$ resists input-output inference with unbounded uncertainty for any $K\in\mathbb{N}$. This completes the proof.
\end{pf}

\section{Conclusion}

This paper studies how to securely execute a class of distributed projected gradient-based algorithms. We propose new homomorphic encryption based schemes which can achieve perfect correctness and simultaneously protect each participant's states and coefficients from any other participant. We further study the issue of input-output inference for a class of quadratic joint functions. The correctness and computational efficiency of the proposed algorithms are verified by two case studies of power systems, one on a demand response problem and the other on an optimal power flow problem. An interesting future work is to relax the assumption of temporarily independent attacks (Assumption \ref{asm: one shot}) for Algorithm \ref{algo: ITS DOP}.

\section{Appendix\label{section appendix}}

\subsection{Literature review on homomorphic encryption\label{literature review on HE}}

\begin{table*}[t!]
\renewcommand{\arraystretch}{1}
\caption{Summary of different types of homomorphic encryption schemes}
\centering
\begin{tabular}{|c|c|c|}
\hline
{} & Partially homomorphic & Fully homomorphic\\
\hline
Private key & Key dist. prob. exists; Efficiently implementable & Key dist. prob. exists; Efficiently implementable\\
\hline
Public key & No key dist. prob.; Efficiently implementable & No key dist. prob.; Not efficiently implementable\\
\hline
\end{tabular}
\label{HE table}
\end{table*}

In general, there are two ways to categorize homomorphic encryption schemes. The first way is based on the keys used in the encryption and decryption operations. Roughly speaking, if the same key is used for both encryption and decryption, then the scheme is referred to as a private key homomorphic encryption scheme; if the keys for encryption and decryption are different and it is computationally infeasible to compute the decryption key from the encryption key, then the scheme is referred to as a public key homomorphic encryption scheme.

The second way of categorization is based on the algebraic operations a homomorphic encryption scheme can carry out. If a scheme can only carry out either addition or multiplication operation but cannot simultaneously carry out both, then the scheme is said to be partially homomorphic; if a scheme can simultaneously carry out both addition and multiplication operations, then the scheme is said to be fully homomorphic.

Table \ref{HE table} summarizes the state-of-the-art of the four types of homomorphic encryption schemes. Private key fully homomorphic (and surely partially homomorphic) encryption schemes can be efficiently implemented but suffer from the fundamental key distribution problem: the message sender and receiver need to share a key which must be kept unknown to any third party. In practice, the key distribution is usually achieved via face-to-face meeting, use of a trusted courier, or sending the key through an existing secure channel. On the other hand, public key homomorphic encryption schemes have the significant advantage of resolving the key distribution problem since the encryption key can be publicized. To the best of our knowledge, all current efficiently implementable public key homomorphic encryption schemes are partially homomorphic, e.g., the RSA scheme \cite{RLR-AS-LA:1978} and the ElGamal scheme \cite{TE:1985} are multiplicatively homomorphic, and the Goldwasser--Micali scheme \cite{SG-SM:1982} and the Paillier scheme \cite{PP:1999} are additively homomorphic. The first fully public key homomorphic encryption scheme was developed by Gentry in his seminal work \cite{CG-2009}. However, due to that the Gentry fully homomorphic encryption scheme uses lattice and bootstrapping, its implementation is overwhelmingly time-consuming. Benchmark implementations include: \cite{CG-SH-NPS:2012} reported an AES (Advanced Encryption Standard) encryption on a supercomputer which took 36 hours; in \cite{CG-SH:2011}, bootstrapping ranged from 30 seconds for small setting to 30 minutes for large setting. The current lack of efficient implementations largely limits the applications of the Gentry fully homomorphic encryption scheme \cite{FA-CB-CC-KG-AJ-CAR-MS:2015}.

\subsection{Preliminaries for the Paillier encryption scheme\label{Paillier priliminaries}}

In this subsection, we provide some preliminaries for the Paillier encryption scheme. The results of this subsection can be found in \cite{PP:1999} and \cite{XY-RP-EB:2014}.

Consider a scenario consisting of a message sender, a message receiver and an adversary. The message sender aims to send a private message to the message receiver via an open communication link which is insecure and can be eavesdropped by the adversary. To achieve secure message delivery, the message sender and receiver can adopt some public key encryption scheme. In this paper, we choose the Paillier encryption scheme as our implementing scheme. The standard Paillier encryption scheme consists of key generation, encryption and decryption operations as follows.

$\bullet$ Key generation: The message receiver randomly chooses two large prime numbers $p$ and $q$ such that ${\rm gcd}(pq,(p-1)(q-1))=1$; computes $\alpha=pq$, $\nu={\rm lcm}(p-1,q-1)$; select random integer $\beta\in\mathbb{Z}_{\alpha^2}^*$ such that the modular multiplicative inverse $\pi=(\frac{(\beta^\nu\mod\alpha^2)-1}{\alpha})^{-1}\mod\alpha$ exists, i.e., $\pi\frac{(\beta^\nu\mod\alpha^2)-1}{\alpha}\equiv1\mod \alpha$. The public keys are $(\alpha,\beta)$ and the private keys are $(\nu,\pi)$. The message receiver publicizes $(\alpha,\beta)$ while keeps $(\nu,\pi,p,q)$ private to itself.

$\bullet$ Encryption: To encrypt a plaintext $pt\in\mathbb{Z}_\alpha$, the message sender selects a random number $r\in\mathbb{Z}_\alpha^*$ and computes the ciphertext by the encryption operation $E(\cdot)$ as $ct=E(pt,\alpha,\beta,r)=\beta^{pt}\cdot r^\alpha\mod \alpha^2$. The message sender then sends $ct$ to the message receiver.

$\bullet$ Decryption: To decrypt the ciphertext $ct\in\mathbb{Z}_{\alpha^2}$, the message receiver performs the decryption operation $D(\cdot)$ as $\overline {pt}=D(ct,\alpha,\nu,\pi)=\frac{(ct^\nu \mod \alpha^2)-1}{\alpha}\cdot \pi \mod \alpha$.

In cryptography, the security of most public-key encryption schemes are established under certain mathematical assumptions which state that certain mathematical problems are difficult to solve. Specifically, the semantic security of the Paillier encryption scheme is proved under the decisional composite residuosity assumption (DCRA), stated as follows: Given a composite $C$ and an integer $z$, it is computationally intractable to decide whether $z$ is a $C$-residue modulo $C^2$ or not, i.e., whether there exists $y$ such that $z=y^C\mod C^2$.

The correctness, security and homomorphic properties of the Paillier encryption scheme are provided by the following theorem, whose proof is given in \cite{PP:1999}.

\begin{theorem}
By the Paillier encryption scheme, the following claims hold:

(1) Correctness: $\overline{pt}=pt$.

(2) Semantic security: If the DCRA holds, then the Paillier encryption scheme is semantically secure.

(3) Homomorphic properties:\\
(3-i) Given any $pt_1,\cdots,pt_m\in\mathbb{N}$. If $\sum_{\ell=1}^mpt_\ell\in\mathbb{Z}_\alpha$, then $D(\prod_{\ell=1}^mE(pt_\ell,\alpha,\beta,r_\ell),\alpha,\nu,\pi)=\sum_{\ell=1}^mpt_\ell$.\\
(3-ii) Given any $pt_1,pt_2\in\mathbb{N}$. If $pt_1 pt_2\in\mathbb{Z}_\alpha$, then $D(E(pt_1,\alpha,\beta,r)^{pt_2},\alpha,\nu,\pi)=pt_1 pt_2$.
\hfill\rule{1mm}{2mm}
\label{theorem: Paillier}
\end{theorem}

\subsection{Numerical examples for Algorithm \ref{algo: ITS DOP} and Algorithm \ref{algo: semantically secure DOP}\label{section numerical examples}}

In this subsection, to help the readers better understand Algorithm \ref{algo: ITS DOP} and Algorithm \ref{algo: semantically secure DOP}, we go through their steps using two simple numerical examples.

\subsubsection{A numerical example for Algorithm \ref{algo: ITS DOP}\label{section numerical example algorithm 1}}

Consider the case of two agents and a system operator. The accuracy level is set as $\sigma=2$. 
%i.e., for any real number, the first 2 decimal fraction digits are kept while the remaining are dropped.
Each agent $i\in\{1,2\}$ has a scalar state $x_i\in\mathbb{R}$. The polynomial joint function of agent 1 is $\Phi_1(x_1,x_2)=c_1x_1^2+c_2x_2^2+c_3x_1x_2+c_4x_1+c_5$, where $c_1,c_3,c_4\in\mathbb{R}$ are the private coefficients of agent 1, $c_2,c_3\in\mathbb{R}$ are the private coefficients of agent 2, and $c_5\in\mathbb{R}$ is the private coefficient of the system operator. Notice that $c_3$ is held by both agent 1 and agent 2. Assume that the participants construct a partition of the coefficient set according to Section \ref{section polynomial preliminaries} as $\tilde C_\Phi^1=\{c_1,c_3,c_4\}$, $\tilde C_\Phi^2=\{c_2\}$ and $\tilde C_\Phi^{\rm SO}=\{c_5\}$. Assume $c_1=3.32$, $c_2=-1.53$, $c_3=4.67$, $c_4=-0.28$, $c_5=2.42$, $x_1(0)=0.76$, $x_2(0)=-2.35$, $X_1=\mathbb{R}$ and $\gamma(0)=1$. One can check $\Phi_1(x_1(0),x_2(0))=-12.665213$. We next go through the steps of Algorithm \ref{algo: ITS DOP} to compute $\Phi_1(x_1(0),x_2(0))$ for one iteration.
%For convenience of notation, we drop the time index $k$.

\noindent\fbox{
\parbox{0.908\linewidth}{
1. \textbf{Initialization:} Agent 1 holds $x_1(0)=0.76$ and agent 2 holds $x_2(0)=-2.35$.
}
}

\noindent\fbox{
\parbox{0.908\linewidth}{
2. \textbf{Key agreement:} Agent 1 and agent 2 agree on a key $w$ according to Assumption \ref{asm: key large enough} and keep it secret from the system operator. We have $1+2\times10^{{\rm deg}(\Phi_1^s)\sigma}|\Phi_1(x_1(0),x_2(0))|=1+2\times10^{3\times2}|3.32\times0.76^2-1.53\times(-2.35)^2+4.67\times0.76\times(-2.35)-0.28\times0.76+2.42|=25330427$. Assume that the key is chosen as $w=25400001$.
}
}

\noindent\fbox{
\parbox{0.908\linewidth}{
3. \textbf{Coefficient encryption:} Agent 1 chooses $u_{y_1}=103$, $u_{y_3}=307$ and $u_{y_4}=205$, and encrypts $(c_1,c_3,c_4)$ as $\hat y_1=u_{y_1}w+10^\sigma c_1=2616200435$, $\hat y_3=u_{y_3}w+10^\sigma c_3=7797800774$ and $\hat y_4=u_{y_4}w+10^\sigma c_4=5207000177$, respectively. Agent 1 sends $(\hat y_1,\hat y_3,\hat y_4)$ to the system operator. Agent 2 chooses $u_{y_2}=501$ and encrypts $c_2$ as $\hat y_2=u_{y_2}w+10^\sigma c_2=12725400348$. Agent 2 sends $\hat y_2$ to the system operator. The system operator computes $\hat y_5=10^\sigma c_5=242$.
}
}

In the above step of coefficient encryption, the security of $(c_1,c_2,c_3,c_4)$ against the system operator is based on the approximate GCD problem, whose definition is given in footnote 1 in page 8. The approximate GCD problem is widely believed to be NP-hard. In the above encryption, the system operator receives $(\hat y_1,\hat y_2,\hat y_3,\hat y_4)$ from the agents, where $\hat y_\ell=u_{y_\ell}w+10^\sigma c_\ell$ for each $\ell\in\{1,2,3,4\}$. To infer $(c_1,c_2,c_3,c_4)$ from $(\hat y_1,\hat y_2,\hat y_3,\hat y_4)$, the system operator has to recover the value of $w$, which is exactly the approximate GCD problem. Hence, it is as hard as the approximate GCD problem for the system operator to solve $(c_1,c_2,c_3,c_4)$.

\noindent\fbox{
\parbox{0.908\linewidth}{
4. Consider the time instant $k=0$ (we only go through one iteration).
}
}

\noindent\fbox{
\parbox{0.908\linewidth}{
5. \textbf{State encryption:} Agent 1 chooses $u_{x_1}(0)=107$, encrypts $x_1(0)$ as $\hat x_1(0)=u_{x_1}(0)w+10^\sigma x_1(0)=2717800183$, and sends $\hat x_1(0)$ to the system operator. Agent 2 chooses $u_{x_2}(0)=409$, encrypts $x_2(0)$ as $\hat x_2(0)=u_{x_2}(0)w+10^\sigma x_2(0)=10388600174$, and sends $\hat x_2(0)$ to the system operator.
}
}

Similar to the security of coefficients discussed above, for the step of state encryption, it is as hard as the approximate GCD problem for the system operator to solve $(x_1(0),x_2(0))$ from $(\hat x_1(0),\hat x_2(0))$.

\noindent\fbox{
\parbox{0.908\linewidth}{
6. \textbf{Computation over ciphertexts:} The system operator computes
\begin{align*}
&\bar\Phi_1^s(0)=\sum_{v=1}^5[10^{({\rm deg}(\Phi_1^s)-{\rm deg}(Q_1^v))\sigma}Q_1^v(\hat x(k),\hat y)]\\
&=10^{(3-3)\times2}\cdot\hat y_1\hat x_1(0)^2+10^{(3-3)\times2}\cdot\hat y_2\hat x_2(0)^2\\
&\,\quad+10^{(3-3)\times2}\cdot\hat y_3\hat x_1(0)\hat x_2(0)\\
&\,\quad+10^{(3-2)\times2}\cdot\hat y_4\hat x_1(0)+10^{(3-1)\times2}\cdot\hat y_5\\
&=1612852152286627752945361608571.
\end{align*}
The system operator sends $\bar\Phi_1^s(0)$ to agent 1.
}
}

In the above step of computation over ciphertexts, we multiply each monomial $Q_1^v$ by $10^{({\rm deg}(\Phi_1^s)-{\rm deg}(Q_1^v))\sigma}$ to addressed the challenge of sum operation over transformed integers as mentioned in item (ii) of Section \ref{section private example}. With the scaling terms, all the monomials are scaled by (the same) $10^6$ times and the sum operation between them can be applied.

\noindent\fbox{
\parbox{0.908\linewidth}{
7. \textbf{Decryption:} Agent 1 computes $\hat\Phi_1^s(0)=T_{{\rm deg}(\Phi_1^s)\sigma,w}(\bar\Phi_1^s(0)\mod w)$. First, we compute $\bar\Phi_1^s(0)\mod w=12734788$. We then have $12700001=(w+1)/2\leq\bar\Phi_1^s(0)\mod w=12734788<w=25400001$. Hence, by \eqref{integer to real}, we have
\begin{align*}
&\hat\Phi_1^s(0)=T_{{\rm deg}(\Phi_1^s)\sigma,w}(\bar\Phi_1^s(0)\mod w)\\
&=\frac{(\bar\Phi_1^s(0)\mod w)-w}{10^{{\rm deg}(\Phi_1^s)\sigma}}=\frac{12734788-25400001}{10^{3\times2}}\\
&=\frac{-12665213}{10^6}=-12.665213.
\end{align*}
}
}

The above result of decryption verifies the perfect correctness of Algorithm \ref{algo: ITS DOP}, i.e., $\hat\Phi_1^s(0)=\Phi_1(x_1(0),x_2(0))=-12.665213$.
%Since $\sigma=2$, we round $\hat\Phi_1^s(0)$ as $\hat\Phi_1^s(0)\approx-12.66$.

\noindent\fbox{
\parbox{0.908\linewidth}{
8. \textbf{Local update:} Agent 1 updates $x_1$ by $x_1(1)=\mathbb{P}_{X_1}[x_1(0)-\gamma(0)\hat\Phi_1^s(0)]=13.425213$.
}
}

\subsubsection{A numerical example for Algorithm \ref{algo: semantically secure DOP}\label{section numerical example algorithm 2}}

Consider the case of two agents and a system operator. The accuracy level is set as $\sigma=2$. Each agent $i\in\{1,2\}$ has a scalar state $x_i\in\mathbb{R}$. The affine joint function of agent 1 is $\Phi_1(x_1,x_2)=A_1^1x_1+A_1^2x_2+B_1$, where $A_1^1,A_1^2,B_1\in\mathbb{R}$ are coefficients known to the system operator. Assume $A_1^1=2.45$, $A_1^2=-3.03$, $B_1=5.22$, $x_1(0)=1.36$, $x_2(0)=-1.42$, $X_1=\mathbb{R}$ and $\gamma(0)=1$. One can check $\Phi_1(x_1(0),x_2(0))=12.8546$. We next go through the steps of Algorithm \ref{algo: semantically secure DOP} to compute $\Phi_1(x_1(0),x_2(0))$ for one iteration.

\noindent\fbox{
\parbox{0.908\linewidth}{
1. \textbf{Initialization:} Agent 1 holds $x_1(0)=1.36$ and agent 2 holds $x_2(0)=-1.42$.
}
}

\noindent\fbox{
\parbox{0.908\linewidth}{
2. \textbf{Key generation:} Agent 1 generates the Paillier keys $(\alpha_1,\beta_1,\nu_1,\pi_1)$ such that $\alpha_1$ satisfies Assumption \ref{asm: public key large enough}. The procedure of Paillier key generation is given in Section \ref{Paillier priliminaries}. We have $1+2\times10^{2\sigma}|A_1^1x_1(0)+A_1^2x_2(0)+B_1|=1+2\times10^{2\times2}|2.45\times1.36+(-3.03)\times(-1.42)+5.22|=257093$. Assume that agent 1 chooses two prime numbers $p_1=733$ and $q_1=523$. Then, we have $\alpha_1=p_1q_1=383359>257093$ and $\nu_1={\rm lcm}(p_1-1,q_1-1)=63684$. Agent 1 chooses $\beta_1=\alpha_1+1=383360\in\mathbb{Z}_{\alpha_1^2}^*$. We then have $\pi_1=(\frac{(\beta_1^{\nu_1}\mod\alpha_1^2)-1}{\alpha_1})^{-1}\mod\alpha_1=198247$. Agent 1 publicizes $(\alpha_1,\beta_1)$ while keeps $(\nu_1,\pi_1,p_1,q_1)$ private to itself.}}

\noindent\fbox{
\parbox{0.908\linewidth}{
3. Consider the time instant $k=0$ (we only go through one iteration).}}

\noindent\fbox{
\parbox{0.908\linewidth}{
4. \textbf{State encryption:} Agent 1 selects $r_1(0)=196827\in\mathbb{Z}_{\alpha_1}^*$, encrypts $x_1(0)$ as $\hat x_1(0)=\beta_1^{10^\sigma x_1(0)\mod\alpha_1}\cdot r_1(0)^{\alpha_1}\mod\alpha_1^2=38891374903$, and sends $\hat x_1(0)$ to the system operator. Agent 2 selects $r_2(0)=199762\in\mathbb{Z}_{\alpha_1}^*$, encrypts $x_2(0)$ as $\hat x_2(0)=\beta_1^{10^\sigma x_2(0)\mod\alpha_1}\cdot r_2(0)^{\alpha_1}\mod\alpha_1^2=112847502000$, and sends $\hat x_2(0)$ to the system operator.}}

The above step of state encryption follows the standard Paillier encryption operation. Since the Paillier encryption scheme is semantically secure, so is the above state encryption against the system operator.
%Hence, the security of $(x_1(0),x_2(0))$ against the system operator follows the security of the Paillier encryption scheme, which is based on the decisional composite residuosity assumption (DCRA). The DCRA is defined as: Given a composite $C$ and an integer $z$, it is computationally intractable to decide whether $z$ is a $C$-residue modulo $C^2$ or not, i.e., whether there exists $y$ such that $z=y^C\mod C^2$ (in the submitted (resp. complete) version of the paper, the definition of the DCRA is given in the sixth paragraph of Section 7.1 (resp. 9.2)). Following the work Paillier (1999), if the DCRA holds, then the above encryption is semantically secure against the system operator.

\noindent\fbox{
\parbox{0.908\linewidth}{
5. \textbf{Computation over ciphertexts:} The system operator computes
\begin{align*}
\bar\Phi_1^s(0)&=\beta_1^{10^{2\sigma}B_1\mod\alpha_1}\cdot\hat x_1(0)^{10^\sigma A_1^1\mod\alpha_1}\\
&\quad\cdot\hat x_2(0)^{10^\sigma A_1^2\mod\alpha_1}\mod\alpha_1^2\\
&=125129165734.
\end{align*}
The system operator sends $\bar\Phi_1^s(0)$ to agent 1.}}

\noindent\fbox{
\parbox{0.908\linewidth}{
6. \textbf{Decryption:} Agent 1 computes $\hat\Phi_1^s(0)=T_{2\sigma,\alpha_1}(\frac{(\bar\Phi_1^s(0)^{\nu_1}\mod\alpha_1^2)-1}{\alpha_1}\cdot\pi_1\mod\alpha_1)$. First, we compute $\frac{(\bar\Phi_1^s(0)^{\nu_1}\mod\alpha_1^2)-1}{\alpha_1}\cdot\pi_1\mod\alpha_1=128546$. Hence, $0\leq\frac{(\bar\Phi_1^s(0)^{\nu_1}\mod\alpha_1^2)-1}{\alpha_1}\cdot\pi_1\mod\alpha_1=128546\leq(\alpha_1-1)/2=191679$. By \eqref{integer to real}, we have
\begin{align*}
&\hat\Phi_1^s(0)\\
&=T_{2\sigma,\alpha_1}(\frac{(\bar\Phi_1^s(0)^{\nu_1}\mod\alpha_1^2)-1}{\alpha_1}\cdot\pi_1\mod\alpha_1)\\
&=\frac{\frac{(\bar\Phi_1^s(0)^{\nu_1}\mod\alpha_1^2)-1}{\alpha_1}\cdot\pi_1\mod\alpha_1} {10^{2\sigma}}\\
&=\frac{128546}{10^{2\times2}}=12.8546.
\end{align*}
}}

The above result of decryption verifies the perfect correctness of Algorithm \ref{algo: semantically secure DOP}, i.e., $\hat\Phi_1^s(0)=\Phi_1(x_1(0),x_2(0))=12.8546$.
%Since $\sigma=2$, we round $\hat\Phi_1^s(0)$ as $\hat\Phi_1^s(0)\approx12.85$.

\noindent\fbox{
\parbox{0.908\linewidth}{
7. \textbf{Local update:} Agent 1 updates $x_1$ by $x_1(1)=\mathbb{P}_{X_1}[x_1(0)-\gamma(0)\hat\Phi_1^s(0)]=-11.4946$.}}

\bibliographystyle{agsm-nq}
\bibliography{MZ,alias,efmain,frazzoli,YL}
\end{document}